\documentclass[12pt]{article}
\usepackage{a4}
\usepackage{amsfonts}
\usepackage{amssymb}
\usepackage{lscape}
\usepackage{cite}
\usepackage{epsfig}
\usepackage{multirow}
\usepackage[all]{xy}
\usepackage{amsmath}
\author{}

\newcommand{\drawsquare}[2]{\hbox{%
\rule{#2pt}{#1pt}\hskip-#2pt%  left vertical
\rule{#1pt}{#2pt}\hskip-#1pt%  lower horizontal
\rule[#1pt]{#1pt}{#2pt}}\rule[#1pt]{#2pt}{#2pt}\hskip-#2pt%  upper horizontal
\rule{#2pt}{#1pt}}% right vertical
\newcommand{\Ysymm}{\raisebox{-.5pt}{\drawsquare{6.5}{0.4}}\hskip-0.4pt%
         \raisebox{-.5pt}{\drawsquare{6.5}{0.4}}}%  symmetric second rank
\newcommand{\Yasymm}{\raisebox{-3.5pt}{\drawsquare{6.5}{0.4}}\hskip-6.9pt%
        \raisebox{3pt}{\drawsquare{6.5}{0.4}}}%  antisymmetric second rank

\newcommand{\be}{\begin{equation}}
\newcommand{\ee}{\end{equation}}
\newcommand{\ba}{\begin{array}}
\newcommand{\ea}{\end{array}}
\newcommand{\bea}{\begin{eqnarray}}
\newcommand{\eea}{\end{eqnarray}}

\newcommand{\ov}{\overline}
\def\IR{\relax{\rm I\kern-.18em R}}

\def\IP{\relax{\rm I\kern-.18em P}}
\def\inbar{\vrule height1.5ex width.4pt depth0pt}
\def\IC{\relax\,\hbox{$\inbar\kern-.3em{\rm C}$}}

\def\K3{{\bf K3}}

\def\ov{\overline}

\def\n2d{\cN_{V^*}^{\otimes 2}}

\def\IC{\mathbb{C}}

\def\IR{\mathbb{R}}

\def\IP{\mathbb{P}}

\def\cN{{\mathcal N}}

\begin{document}

\title{
\begin{flushright} \vspace{-2cm}
{\small UPR-1208-T\\
%\vspace{-0.35cm}
} \end{flushright} \vspace{4.0cm}
Realistic Yukawa Structures from Orientifold Compactifications }
 \vspace{1.0cm}
\author{\small  Mirjam Cveti{\v c}, James Halverson,  Robert Richter}

\date{}

\maketitle

\begin{center}
\emph{Department of Physics and Astronomy, University of Pennsylvania, \\
     Philadelphia, PA 19104-6396, USA }
\vspace{0.2cm}

\tt{cvetic@cvetic.hep.upenn.edu, jhal@physics.upenn.edu, rrichter@physics.upenn.edu }
\vspace{1.0cm}
\end{center}
\vspace{0.7cm}

%%%% new

\begin{abstract}
We perform a systematic analysis of globally consistent D-brane quivers which realize the MSSM and analyze them with respect to their Yukawa couplings.
Often, desired couplings are perturbatively forbidden due to the presence of global $U(1)$ symmetries. We investigate the conditions
under which D-brane instantons will induce these missing couplings without generating other phenomenological drawbacks, such as R-parity violating
couplings or a $\mu$-term which is too large. Furthermore, we systematically analyze which quivers
allow for a mechanism that can account for the small neutrino masses and other experimentally observed hierarchies.
We show that only a small fraction of the globally consistent D-brane quivers exhibits phenomenology compatible with experimental observations.
\end{abstract}
 
\thispagestyle{empty} \clearpage

%\tableofcontents

%\center{
%{\Large\textbf{Paperv15}}
%\\
\section{Introduction}

Intersecting D-brane models have been proven to be a fruitful playground for realistic model building (for reviews on this subject see \cite{Blumenhagen:2005mu, Marchesano:2007de,Blumenhagen:2006ci}). In these string compactifications, the gauge groups arise from stacks of D6-branes that fill out four-dimensional
spacetime and wrap three-cycles in the internal Calabi-Yau threefold. Chiral matter appears at the intersection in the internal space of different cycles wrapped by the D6-brane stacks. The multiplicity of chiral matter between two stacks of D6-branes is given by the topological intersection number of the respective three-cycles.

Over the last decade, many semi-realistic MSSM-like and supersymmetric GUT-like realizations have been constructed based on intersecting branes,
mostly using toroidal orbifold compactifications\footnote{For
original work on non-supersymmetric intersecting D-branes, see
\cite{Blumenhagen:2000wh,Aldazabal:2000dg,Aldazabal:2000cn,Blumenhagen:2001te},
and for chiral supersymmetric ones, see
\cite{Cvetic:2001tj,Cvetic:2001nr}.  For supersymmetric MSSM realizations, see \cite{Honecker:2004kb,Gmeiner:2007zz,Gmeiner:2008xq} and for supersymmetric MSSM-like constructions within type II RCFT's, see \cite{Dijkstra:2004cc,Anastasopoulos:2006da}.}.
Once the spectrum of a particular string compactification is determined, finer details
can be investigated, such as the Yukawa couplings for the chiral matter fields.  In intersecting
 brane compactifications such Yukawa couplings can be extracted from string amplitudes
 \cite{Cvetic:2003ch,Abel:2003vv,Lust:2004cx,Bertolini:2005qh}. These amplitudes are suppressed by open string
 world-sheet instantons connecting the three intersecting branes \cite{Aldazabal:2000cn,Cremades:2003qj}
 and thus could potentially give rise to hierarchies observed in nature.
However, often times these desired couplings are forbidden due to the violation of global U(1)
symmetries\footnote{As demonstrated in
\cite{Cvetic:2002wh} (see also
\cite{Chen:2007px,Chen:2007zu,Chen:2008rx,Anastasopoulos:2009mr}) allowing for additional Higgs pairs
might give rise to all Yukawa couplings, and, for appropriate values of the open
and closed string moduli, give rise to desired Yukawa hierarchies.}. The latter are remnants of the Green-Schwarz mechanism, which is crucial for the cancellation of abelian, mixed and gravitational anomalies.

Recently, it has been realized that D-brane instantons  can break these global symmetries and  generate
otherwise forbidden coupings\cite{Blumenhagen:2006xt,Haack:2006cy,Ibanez:2006da,Florea:2006si}\footnote{For a recent review on the D-instanton effects see \cite{Blumenhagen:2009qh} and also \cite{Akerblom:2007nh,Cvetic:2007sj}.}. For type IIA compactifications the relevant objects are so called $E2$-instantons,
which wrap a three-cycle in the internal manifold and are point-like in space-time.
They are charged under the global U(1)'s, where the charge is given by\footnote{Note that, in contrast to \cite{Blumenhagen:2006xt}, there
is an additional minus sign in \eqref{eq charge generic instanton},
which is due to the fact that a positive intersection number
$I_{E2a}$ corresponds to the transformation behavior
$(E2,\overline{a})$ of the charged fermionic zero modes, rather than
$(\overline{E2},a)$.}
\begin{equation}
Q_a(E2)=-N_a\, \pi_{E2} \circ (\pi_a+\pi'_a)\,\,.
\label{eq charge generic instanton}
\end{equation}
Here the subscript $a$ refers to the global $U(1)_a$ arising from the stack of $N_a$ D6 branes which wrap the cycle $\pi_a$. The instanton itself wraps the cycle $\pi_{E2}$ and $\pi'_a$ denotes the orientifold image cycle of $\pi_a$.

A superpotential term which violates global $U(1)$ symmetries is perturbatively forbidden, but can be generated non-perturbatively
\begin{equation}
W^{np}=e^{-S_{E2}} \prod_{i} \Phi_{a_i\,b_i}\,\,,
\end{equation}
if the  $E2$ instanton compensates for the global U(1) charge carried by the product $\prod_{i} \Phi_{a_i\,b_i}$. The  exponential suppression factor due to the classical instanton action depends on the volume of the three-cycle wrapped by the instanton and is given by
\begin{equation}
e^{-S^{cl}_{E2}}=e^{-\frac{2 \pi}{l^3_s g_s} Vol_{E2}}
\,\,.
\label{eq suppression factor}
\end{equation}
Note that the instanton is independent of the 4D gauge couplings and therefore has no interpretation as a gauge instanton. Rather, it is purely stringy. Due to this property, D-instantons not only induce perturbatively forbidden couplings but also may give a natural explanation for hierarchies, which are poorly understood from a field theoretical point of view.

Apart from the charged instanton modes, which arise from strings attached to the instanton and a D6-brane, a generic D-instanton also exhibits the so called uncharged instanton modes.
The latter consist of the four bosonic modes $x^{\mu}$, which are associated with breakdown
of four-dimensional Poincar{\'e} invariance, and the fermionic modes $\theta^{\alpha}$ and
$\overline{\tau}^{\dot{\alpha}}$, which indicate the breakdown of the ${
\cal N}=2 $ supersymmetry preserved by the Calabi-Yau manifold
down to ${\cal N}=1$ supersymmetry. Additional zero modes appear if the instanton wraps a non-rigid three-cycle in the internal manifold. In the presence of multiple instantons, there may arise zero modes at intersections of
two instantons.

The non-perturbative contribution to the superpotential is given by the path integral over
all instanton zero modes. Thus, in order to give rise to superpotential terms, one has to ensure
that all uncharged zero modes, apart from $x^{\mu}$ and
$\theta^{\alpha}$, are projected out or lifted. There are various
mechanisms to ensure the saturation of these additional undesired zero modes,
such as lifting via
fluxes \cite{Blumenhagen:2007bn,Billo':2008sp,Billo':2008pg,Uranga:2008nh},
saturation via multi-instanton configurations
\cite{Blumenhagen:2007bn,GarciaEtxebarria:2007zv,Cvetic:2008ws,GarciaEtxebarria:2008pi},
or via additional interaction terms arising when the instanton wraps a
cycle which coincides with one of the spacetime filling D6-branes
\cite{Akerblom:2006hx,Billo:2007py,Petersson:2007sc,Ferretti:2009tz}. 

If the E2-instanton wraps a rigid orientifold invariant cycle, the zero modes are subject to the orientifold projection and
the $\ov \tau_{\dot{\alpha}}$ modes are projected out \cite{Argurio:2007qk,Argurio:2007vq,
Bianchi:2007wy,Ibanez:2007rs}. This is referred to as a rigid $O(1)$ instanton and can contribute to the superpotential, due to the
absence of the $\ov \tau_{\dot{\alpha}}$ modes. For simplicity and clarity in this analysis,
we will assume that rigid $O(1)$ instantons generate the missing couplings.
For an $O(1)$ instanton  the $E2a$ and $E2a'$ sector are identified, which modifies the charge of the instanton under $U(1)_a$ to
\begin{equation}
Q_a(E2)=-N_a\, \pi_{E2} \circ \pi_a\,\,.
\label{eq U(1)-charge of O(1) instanton}
\end{equation}
We emphasize, though, that any other instanton configuration with the same charged zero mode structure is equally good and the analysis performed applies analogously\footnote{This holds not true for instanton configurations where zero modes are lifted via fluxes. The latter contribute to the tadpole equations and thus modify the analysis.}.

It has been explicitly shown that D-instantons carrying the correct zero mode structure can generate
Majorana masses for right-handed neutrinos \cite{Blumenhagen:2006xt,Ibanez:2006da,Cvetic:2007ku,Ibanez:2007rs,Antusch:2007jd,Cvetic:2007qj,Kiritsis:2008ry},
induce a $\mu$ term for the Higgs pair \cite{Blumenhagen:2006xt,Ibanez:2006da,Ibanez:2007rs}, and
generate the $\mathbf{10\,10 \,5}$ Yukawa coupling in $SU(5)$-GUT-like models \cite{Blumenhagen:2007zk,Kokorelis:2008ce}.
Moreover, these nonperturbative effects are relevant for supersymmetry breaking
\cite{Aharony:2007db,Buican:2008qe,Cvetic:2008mh,Heckman:2008es,Marsano:2008jq,Marsano:2008py,Bianchi:2009bg}
and moduli stabilization \cite{Kachru:2003aw,Denef:2004dm,Balasubramanian:2005zx,Denef:2005mm,
Lust:2005dy,Lust:2006zg,Blumenhagen:2007sm,Palti:2008mg,Blumenhagen:2008kq}.

In \cite{Ibanez:2008my} the authors analyzed specific
semi-realistic four-stack D-brane quivers in which all of the matter content transforms as bifundamentals
 of the respective gauge groups, and thus these setups do not exhibit any matter transforming as symmetric
 or antisymmetrics. For these quivers, they analyzed under what circumstances perturbatively missing couplings
 can be generated via D-instantons and investigated their phenomenological implications.
 A potential problem for this class of D-brane quivers is that 
the instantons which induce the Yukawa couplings lead to phenomenologically undesirable effects, such as a
$\mu$-term which is too large or too much family mixing. As also shown in \cite{Ibanez:2008my},
these phenomenological drawbacks can be circumvented if one allows for a second Higgs pair.

In this work we follow a slightly different path.
Here, instead of looking at concrete semi-realistic realizations of the standard model,
we focus on the whole class of all globally consistent D-brane quivers which exhibit the standard model
gauge symmetry and the exact MSSM matter spectrum plus three right-handed neutrinos \cite{Anastasopoulos:2006da}.
We investigate these quivers with respect to their Yukawa couplings and analyze in each case whether instantons
can generate perturbatively forbidden, but desired, couplings without inducing tadpoles
or R-parity violating couplings. In addition, we require that these quivers exhibit a
mechanism which accounts for the smallness of the neutrino masses, give rise to a small
$\mu$-term, and give Yukawa textures compatible with experimental observations without too much fine-tuning. Let us emphasize that this analysis is completely independent of any concrete global realization and is thus a bottom-up analysis \cite{Antoniadis:2000ena,Aldazabal:2000sa,Antoniadis:2001np}.

We show that only a small subset of the globally consistent D-brane quivers exhibits such phenomenology.
The whole analysis is independent of any concrete realization
and thus the results serve as a good starting point for future model building.
Furthermore, even though this analysis is performed in the type IIA corner
of the string theory landscape, the results also apply to the T-dual type I framework as well as to type IIB with D-branes on singularities.

This paper is organized as follows. In chapter \ref{sec constraints} we describe the criteria
and conditions for our MSSM realizations. We start by investigating the implications of the top-down constraints, which include
tadpole cancellation and presence of a massless $U(1)_Y$, for D-brane quivers and then proceed to discuss a
number of bottom-up constraints, motivated by experimental observations.
In section \ref{sec three-stack models}, we present all solutions to our constraints arising from
three-stack models where the D-branes wrap generic cycles. We discuss phenomenological properties
of each solution in detail, including the role played by instantons. We see that some of these setups
exhibit phenomenological drawbacks, such as too much family mixing or the presence of R-parity violating
couplings, once non-perturbative effects are taken into account.
In section \ref{sec four-stack models} we analyze four stack models, which give rise to a much richer structure.
We embed further bottom-up constraints motivated by the lessons learned in section \ref{sec three-stack models}
and in \cite{Ibanez:2008my}. We present the subset of D-brane quivers compatible with these phenomenological
considerations. This subset represents a good starting point for concrete MSSM realizations with realistic Yukawa structure.
In section \ref{chapsp(2)} we consider D-brane quivers where the $SU(2)_L$ of the MSSM arises
from a stack of D-branes wrapping an orientifold invariant cycle and perform an analysis
similar to those in sections \ref{sec three-stack models} and \ref{sec four-stack models}.
We conclude in section \ref{sec conclusion} with a summary of our results and a brief outlook.

\section{Constraints \label{sec constraints}}
In this chapter we discuss criteria and conditions that we require of the MSSM realizations.
Some conditions, such as tadpole cancellation and the presence of a massless hypercharge,
are top-down constraints. Others, such as conditions on the spectrum and
on the superpotential, are bottom-up constraints.

Before discussing these criteria and conditions in detail let us
describe the generic setup of MSSM quivers based on three and four
stacks of D-branes. The standard model matter content arises at
intersections of three or four stacks of D-branes which give rise to the gauge
symmetry\footnote{The $SU(2)$ can be also realized as $Sp(2)$. This case will be discussed in chapter \ref{chapsp(2)}. }
\begin{align*}
U(3)_a\times U(2)_b \times U(1)_c \qquad
U(3)_a\times U(2)_b \times U(1)_c \times U(1)_d \,\,.
\end{align*}
The tadpole conditions imply the vanishing of non-abelian anomalies, while
abelian and mixed anomalies are cancelled via the Green-Schwarz
mechanism. Generically, the anomalous $U(1)$'s acquire a mass and
survive only as global symmetries, which forbid various couplings on the perturbative level.
Since the standard model gauge symmetries contain the abelian symmetry $U(1)_Y$, we require that a
linear combination
\begin{align}
U(1)_Y= \sum_{x} q_x \,U(1)_x \,\,,\label{eq hypercharge}
\end{align}
remains massless. Thus, the
resulting gauge group in four-dimensional spacetime is
\begin{align*}
SU(3)_a \times SU(2)_b \times U(1)_Y\,\,.
\end{align*}

Let us briefly comment on the origin of the MSSM spectrum. The
left-handed quarks $q_{L}$ are localized at intersections of brane
$a$ and $b$ or its orientifold image $b'$, while the right-handed
quarks, $u_R$ and $d_R$, arise at intersections of brane $a$ with
one of the $U(1)$ branes or its orientifold image. Depending on the
hypercharge, the right-handed quarks can also transform as
antisymmetric of $SU(3)$. The left-handed leptons and the Higgs
fields, $H_u$ and $H_d$, are charged under the $SU(2)$ and neutral
under $SU(3)$, and thus appear at intersections between brane $b$ and
one of the $U(1)$ branes. Finally, the right-handed electrons $E_R$ and neutrinos $N_R$,
which are singlets under $SU(3)$ and $SU(2)$, can arise at intersections of two
$U(1)$ branes or at the intersection $bb'$, in which case they would transform as antisymmetric of $SU(2)$.
We emphasize that the actual origin of the matter fields crucially depends on the choice of hypercharge $U(1)_Y$.

We impose constraints on the Yukawa couplings and neutrino masses
according to experimental observations. Specifically, we require that the
Yukawa couplings $q_L\,H_u\,u_R$, $q_L\,H_d\,d_R$, and $L\,H_d\,E_R$ give rise
to three massive families of u-quarks, d-quarks, and electrons. Furthermore, we forbid the presence of R-parity violating couplings and
we require the presence of a mechanism which accounts for the observed smallness of the neutrino masses.

We start by analyzing the top-down constraints, tadpole
cancellation and the presence of a massless $U(1)$. Both constraints, which are conditions on the cycles the D-branes
wrap, imply restrictions on the transformation properties of the chiral
spectrum. This will be discussed in the sections \ref{sec tadpole} and \ref{sec massless U(1)}.
The bottom-up constraints, which contain conditions on the spectrum and superpotential, will be subject in sections
\ref{sec spectrum}, \ref{sec Yukawa} and \ref{sec neutrino masses}.
Specifically, we require that the chiral spectrum arising from the three or four
stacks reproduces the MSSM spectrum plus three right-handed neutrinos and does not allow for any additional chiral exotics.
In \ref{sec spectrum} we define precisely what we mean by the latter. In section \ref{sec Yukawa}
we discuss the constraints arising from the MSSM superpotential and in section \ref{sec neutrino masses} we present two mechanisms which can account for
small neutrino masses.

\subsection{Tadpole Condition \label{sec tadpole}} The tadpole
condition is a constraint on the cycles the D-branes wrap and reads
in type IIA
\begin{align*}
\sum_{x} N_x \left(\pi_x +\pi'_{x} \right) -4\pi_{O6}=0\,\,.
\end{align*}
Multiplying this equation with the homology class of the cycle
that is wrapped by a stack $a$ gives, after a few manipulations
\begin{align*}
\sum_{x\neq a} N_{x} (  \pi_x \circ
\pi'_a-\pi_{x} \circ \pi_{a})+\frac{N_a-4}{2} (\pi_a \circ \pi'_a+ \pi_a \circ \pi_{O6}) +
\frac{N_a+4}{2} (\pi_a \circ \pi'_a- \pi_a \circ \pi_{O6})=0\,\,,
\end{align*}
which constrains the transformation behavior of the chiral matter fields.
\begin{table}
\centering
\begin{tabular}{|c|c|}
\hline
Representation  & Multiplicity \\
\hline $ \Yasymm_a$
%$[{\bf A_a}]_{L}$
 & ${1\over 2}\left(\pi_a\circ \pi'_a+\pi_a \circ  \pi_{{\rm O}6} \right)$  \\
$\Ysymm_a$
%$[{\bf S_a}]_{L}$
     & ${1\over 2}\left(\pi_a\circ \pi'_a-\pi_a \circ  \pi_{{\rm O}6} \right)$   \\
$( a,{\overline b})$
%$[{\bf (\o N_a,N_b)}]_{L}$
 & $\pi_a\circ \pi_{b}$   \\
 $(a,b)$
%$[{\bf (N_a, N_b)}]_{L}$
 & $\pi_a\circ \pi'_{b}$
\\
\hline
\end{tabular}
\vspace{2mm} \caption{Chiral spectrum for intersecting D6-branes.}
\label{table chiral spectrum}
\end{table}
Given the relations displayed in Table \ref{table chiral spectrum},
one gets
\begin{align}
\#(a) - \#({\ov a})  + (N_a-4)\#( \,\Yasymm_a) + (N_a+4) \#
(\Ysymm_a)=0 \,\,,\label{eq constraint1}
\end{align}
where $\#(a)$ gives the total number of fields transforming as
fundamental under $SU(N_a)$ and analogously for $\#({\ov a})$,
$\#(\,\Yasymm_a)$ and $\#(\Ysymm_a)$. Equation \eqref{eq constraint1} is
nothing else than the anomaly cancellation for non-abelian gauge
theories for $SU(N_a)$ gauge groups with rank $N_a>2$.
 It holds still
true for $N_a=2$, i.e $SU(2)$, where $2 = {\ov 2}$, but note
that string theory distinguishes between $2$ and ${\ov 2}$.
%In the absence of any symmetric and antisymmetric fields the
%tadpole constraints imply the number of fundamentals is equal to number of antifundamentals.

Let us turn to the case $N_a=1$. The $U(1)$ does not give rise to any
antisymmetrics, thus for $N_a=1$ the constraint \eqref{eq constraint1}
reduces to
\begin{align}
\#(a) - \#({\ov a}) + 5 \# (\Ysymm_a)=0  \qquad \text{mod} \,3\,\,.
\label{eq constraint2}
\end{align}

Summarizing, tadpole cancellation implies constraints on the transformation properties of the chiral matter fields.
For $N_a>2$ it is the usual non-abelian anomaly cancellation. Since the chiral spectrum is the exact MSSM
plus three right-handed neutrinos the constraint for $U(3)$ is trivially satisfied.
On the other hand for $U(2)$ and the $U(1)$ tadpole cancellation puts non-trivial
constraints, given by \eqref{eq constraint1} and \eqref{eq constraint2}, on the transformation behavior of the matter fields.

\subsection{Massless U(1)'s
\label{sec massless U(1)}}
Generically, the $U(1)$ gauge bosons acquire a mass term via the
Green-Schwarz mechanism, which ensures the cancellation of pure
abelian, mixed and gravitational anomalies. The massive $U(1)$'s are not part of the low
energy effective gauge symmetry, but remain as unbroken global symmetries  at
 perturbative level and thus may forbid  various desired couplings.

A linear combination 
\begin{align}
U(1) = \sum_{x} q_{x} \,U(1)_{x}
\label{eq massless U(1)}
\end{align}
remains massless if the condition\cite{Aldazabal:2000dg}
\begin{align}
\sum_{x} q_{x}\,N_{x} (\pi_{x} -\pi'_{x}) =0
\label{eq cond for masllessness}
\end{align}
is satisfied. Here $x$ denotes the different D-brane stacks present
in the model. Since the standard model contains the $U(1)_Y$ hypercharge as a
gauge symmetry, we require the presence of massless $U(1)$ which can be identified with $U(1)_Y$.
As we will see in chapter \ref{sec MSSM} only for a few combinations \eqref{eq massless U(1)} do
all the matter particles have the proper hypercharge \cite{Anastasopoulos:2006da}.

Analogously to the analysis performed for the tadpole constraints we multiply
equation \eqref{eq cond for masllessness} with the homology class $\pi_a$  of the
cycle wrapped by the D-brane stack $a$. After a few manipulations and applying the
relation given in Table \ref{table chiral spectrum}, we obtain
\begin{align}
- q_a\,N_a \,\Big(\#(\Ysymm_a) + \# (\,\Yasymm_a)\Big) + \sum_{x \neq a} q_x\,
N_x \#(a,{\ov x}) - \sum_{x \neq a} q_x\, N_x \#(a,x)=0\,\,.
\label{eq massless constraint non-abelian}\end{align}
Note that \eqref{eq massless constraint non-abelian} gives a constraint for every D-brane stack present in the model.
Thus for three and four stack models we expect three and four additional constraints respectively, due to the presence of a massless hypercharge.

Due to the absence of antisymmetric matter for $U(1)$'s the condition \eqref{eq massless constraint non-abelian}
applies only to nonabelian gauge symmetry. For abelian gauge groups, the condition reads
\begin{align}
-q_a\,\frac{\#(a) - \#({\ov a}) + 8 \#
(\Ysymm_a)}{3}  + \sum_{x \neq a} q_x\, N_x \#(a,{\ov x}) -
\sum_{x \neq a} q_x\, N_x \#(a,x)=0\,\,.
\label{eq massless constraint abelian}
\end{align}

\subsection{Chiral and Non-chiral Spectrum
\label{sec spectrum}}
Here we discuss the origin of the MSSM spectrum and give a precise definition of chiral
exotics. In order to do so, let us split the whole class of D-brane stacks into two
disjoint classes $O$ and $H$. Here $O$, the observable class, corresponds to the
three or four D-brane stacks and their orientifold images from which
the MSSM matter arises. Generically, additional
hidden D-brane stacks are present to ensure the cancellation of the RR-charges.
The latter are elements of the subclass $H$.

In this work we require that all MSSM fields plus the three
right-handed neutrino $N_R$  are only charged under the subclass
$O$\footnote{Note that this is in contrast to the analysis of \cite{Anastasopoulos:2006da},
where the authors allowed for right-handed neutrinos from the hidden sector.}.
Thus we do not allow for any MSSM matter
field to appear at intersections between a brane of class $O$ and a hidden
brane of class $H$. Moreover, we forbid any MSSM matter fields arising
from two D-branes of type $H$, and we require the absence of
any chiral fields charged under any D-brane stacks in the observable sector in addition to the MSSM spectrum.
Thus all matter fields charged under $O$ and $H$ appear as
vector-like state and potentially receive a large mass once the open
string moduli are fixed. We emphasize again that the $OO$ sector contains only
 MSSM matter fields, so that chiral exotics can only appear within the hidden sector.

These assumptions require, as an immediate consequence, that the
constraints on the transformation behavior arising from the tadpole
cancellation as well as from the appearance of a massless
hypercharge $U(1)_Y$ have to be satisfied within the observable class
$O$. This is due to the fact that non-chiral matter does not enter the
constraints derived in sections \ref{sec tadpole} and \ref{sec massless U(1)}.
Thus $x$ in the constraints \eqref{eq constraint1},\eqref{eq constraint2},
\eqref{eq massless constraint non-abelian} and \eqref{eq massless constraint abelian}
is an element of $O$ and runs from $1$ to the number of stacks.

\subsection{Yukawa Couplings \label{sec Yukawa}}

The superpotential of the MSSM contains the terms
\begin{align}
q_L \, H_u \, u_R  \qquad q_L\, H_d \,d_R \qquad  L \,H_d \,E_R \qquad  L \,H_u \,N_R  \qquad  H_u \, H_d \,\,.
\label{eq MSSM couplings}
\end{align}
Any realistic string vacua has to exhibit such terms in its superpotential. If the smallness
of the neutrino masses is due to the type I seesaw mechanism, then the presence of a mass term
\begin{align}
N_R\, N_R
\label{majorana mass term}
\end{align}
for the right-handed neutrinos is required. All these couplings have to be realized either perturbatively or via D-instanton
effects, as discussed in the introduction. The MSSM allows for additional couplings
\begin{align}
u_R \, d_R \, d_R \qquad q_L\, L \,d_R \qquad  L\,L\,E_R  \qquad L\,H_u
\label{eq R-parity violating couplings}
\end{align}
which are invariant under the standard model gauge symmetries. These are referred to as R-parity violating
couplings and could lead to a rapid proton decay, which is in contradiction with experiments. In this
work we require the absence of any of these couplings, both perturbatively and non-perturbatively.
Let us mention that the experimental constraints on just $L$- or $B$-number violating couplings are not
as stringent, as long as it is ensured that the setup preserves one of the two symmetries. Nevertheless,
such couplings have to be suppressed compared to the MSSM Yukawa couplings \eqref{eq MSSM couplings}.
This implies that R-parity violating couplings should be perturbatively absent. Moreover, if some of
the couplings in \eqref{eq MSSM couplings} are perturbatively forbidden, and thus have to be induced via
instantons, we require the very same instanton which induces the desired Yukawa coupling does not
generate any R-parity violating couplings \eqref{eq R-parity violating couplings}. Otherwise
the latter are expected to be of the same order as the instanton induced Yukawa coupling, which
is not compatible with experimental observations. Also, in the absence of R-parity violating couplings,
the lightest supersymmetric particle cannot decay and could therefore be a dark matter candidate.

Let us briefly comment on R-parity violating couplings involving the right-handed neutrinos.
The Majorana mass term \eqref{majorana mass term} violates R-parity, but its presence does not
lead to any contradictions with experiments. The other gauge invariant couplings,
\begin{align}
N_R\, N_R\, N_R \qquad H_u \,H_d \,N_R
\label{eq R-parity involving N_R}
\end{align}
which are R-parity violating, may affect the Higgs potential and thus the Higgs VEV. We do not forbid the presence of such terms.

Finally, we require that instantons required to generate Yukawa couplings do not also induce terms of the form
\begin{align}
N_R
\end{align}
in the superpotential.
The latter is a tadpole and would indicate an instability of the string vacuum.
\subsection{Neutrino Masses
\label{sec neutrino masses}}
Finally, let us discuss the neutrino masses. Experimental observations indicate that the neutrino mass
is very small compared to the masses of any other chiral matter.
Here we allow for two different mechanisms which explain this hierarchy. The first
one is the well known Type I seesaw mechanism. A necessary ingredient for this mechanism
is the presence of a Majorana mass term for the right-handed neutrinos, in addition to the
usual Dirac mass term. The second scenario is string inspired \cite{Cvetic:2008hi} and assumes
that the Dirac mass term is perturbatively forbidden. An instanton is used to induce this term
with high suppression, and thus explains the hierarchy between neutrino mass and all other standard
model particle masses. For various quivers we will also encounter hybrids of these two mechanisms.

Let us briefly comment on the Weinberg operator, which provides another potential
explanation for the smallness of the neutrino masses. Such an operator is generically
perturbatively forbidden, but could in principle be generated via D-instantons, giving rise to
a term
\begin{align}
e^{-S_{E2}} \frac{L \, H_u \, L \,H_u }{M_s}\,\,.
\label{eq weinberg operator}
\end{align}
Assuming that $M_s$ is of the order $10^{18}$ GeV and the Higgs VEV is
around $100 \,GeV$ leads to neutrino masses which are too small, even with a negligible
suppression factor\footnote{In principle the string mass can be lower if we allow
for large extra dimensions.} \cite{Ibanez:2007rs}.

\section{MSSM Quivers and Their Phenomenology}
\label{sec MSSM}
In this chapter, we will study the three and four stack models
which give rise to the exact MSSM plus three right-handed
neutrinos. As we discussed in section \ref{sec tadpole}, this puts severe
constraints on the transformation behavior of the matter
fields. Matching the MSSM also requires a massless hypercharge,
which can only occur if equations \eqref{eq massless constraint non-abelian} and
\eqref{eq massless constraint abelian} are satisfied.

We study setups which satisfy these constraints with respect to their
Yukawa couplings. To be more precise, we
investigate which Yukawa couplings are perturbatively realized and
under what circumstances non-perturbative effects can induce
perturbatively forbidden couplings. Furthermore, we analyze the
setups with respect to R-parity violating couplings and investigate
if the setups allow for a $\mu$-term and neutrino masses of the
observed order.

We start by investigating quivers based on three stacks of D-branes. For these we will
discuss in detail all setups which match the MSSM spectrum, satisfy the constraints
\eqref{eq constraint1}, \eqref{eq constraint2}, \eqref{eq massless constraint non-abelian} and \eqref{eq massless constraint abelian}, and do
not give rise to R-parity violating couplings on the perturbative
level. We will investigate under what circumstances
D-brane instantons can generate perturbatively forbidden, but
desired, couplings and analyze further phenomenological implications
of the presence of such instantons. As we will see, these
additional effects cause some of the quivers to be unrealistic.

In section \ref{sec four-stack models} we allow for an additional $U(1)$ D-brane stack.
This enlarges the number globally consistent D-brane quivers.
We perform a systematic analysis for all these quiver with respect to the constraints studied in chapter \ref{sec constraints} and
embed further bottom-up conditions motivated by the lessons learned in section \ref{sec three-stack models}.
All constraints are summarized in detail at the beginning of section \ref{sec four-stack models}. It
turns out that only a small subset of the globally consistent D-brane quivers exhibit a desirable
phenomenology. For each choice of hypercharge we list these quivers, which serve as starting point for future model building explorations.

Later in section \ref{chapsp(2)} we study quivers where the $SU(2)_L$ is realized as  an $Sp(2)$ gauge group.
Sticking to the exact MSSM and 3 right-handed neutrinos, it is easy to prove that there exists only one
quiver which does not give rise to R-parity violating couplings. This particular quiver was locally
realized and analyzed in \cite{Cremades:2002va,Cremades:2003qj}. All the MSSM Yukawa couplings are
perurbatively present and we show that a D-instanton which induces a large Majorana mass term can
account for small neutrino masses.

\subsection{Three-stack Models \label{sec three-stack models}}

The most economical way to realize the MSSM spectrum is to use three
stacks of D-branes wrapping generic cycles, which give rise to the gauge symmetry $U(3)_a
\times U(2)_b \times U(1)_c$ in four-dimensional space-time.
Generically, the abelian parts of $U(3)_a$ and $U(2)_b$, as well as
$U(1)_c$ itself, are anomalous. The Green-Schwarz
mechanism, which ensures the cancellation of these anomalies, promotes
 the abelian gauge symmetries to global $U(1)$ symmetries which are respected by all perturbative couplings.
In order to match the MSSM gauge symmetry, we require that
a linear combination
\begin{align}
U(1)_Y= q_a \,U(1)_a + q_b\, U(1)_b + q_c\, U(1)_c,
\end{align}
identified as the hypercharge $U(1)_Y$, does not acquire a St\"uckelberg mass term, and thus remains massless.
One finds two different
linear combinations\footnote{Up to minus sign in front of $U(2)_b$ and
$U(1)_c$.} for the hypercharge which are consistent with the MSSM hypercharge
assignments, namely
\begin{align}
U(1)_Y=\frac{1}{6} U(1)_a + \frac{1}{2} U(1)_c \qquad
U(1)_Y=-\frac{1}{3} U(1)_a - \frac{1}{2} U(1)_b\,\,.
\end{align}
In the following we analyze the two different cases. We present all
possible realizations of the MSSM for each choice of hypercharge and
investigate each quiver with respect to its Yukawa couplings.
If they are perturbatively forbidden, we examine under what
conditions they can be generated by D-instantons.

\subsubsection{$U(1)_Y=\frac{1}{6} U(1)_a + \frac{1}{2} U(1)_c$
\label{sec threestack1 }}

We start by examining the first case. For this choice of hypercharge, the right-handed
electron arises at intersections between the brane $c$ and its orientifold image
$c'$. The right-handed neutrino is located at intersections between
$b$ and $b'$ and transforms as antisymmetric under $SU(2)$. The
right-handed d-quarks $d_R$ have two potential origins, since they can
arise from the sector $aa'$ or sector $ac$. Similarly, the left-handed
quarks $q_L$ have two potential origins, coming either from the sector $ab$
or the sector $ab'$. The right-handed u-quarks $u_R$ are localized at intersections between stack $a$ and $c'$.

Below we summarize the potential origins of all the matter
fields. Here the $a$ and $\ov{a}$ correspond to fundamental and
antifundamental representations of the gauge group
$U(3)_a$, and similarly for the other stacks. The Young diagrams $\Ysymm$ and $\Yasymm$
denote fields transforming as symmetric and antisymmetric representations of the respective gauge symmetry.
\begin{align*}
q_L\,:&\qquad (a,\ov{b}), \,\,\,\,\, (a,b)\\
u_R\,:&\qquad (\ov{a},\ov{c})\\
d_R\,:&\qquad (\ov{a},c),\,\,\,\,\, {\Yasymm}_a\\
L\,\,:&\qquad (b,\ov{c}),\,\,\,\,\,(\ov{b},\ov{c})\\
E_R\,:&\qquad \Ysymm_c\\
N_R\,:&\qquad {\Yasymm}_b,\,\,\,\,\, \ov{\Yasymm}_b\\
H_u\,:&\qquad (\ov{b},c), \,\,\,\,\, (b,c)\\
H_d\,:&\qquad (b,\ov{c}), \,\,\,\,\, (\ov{b},\ov{c})\\
\end{align*}
Out of all possible MSSM setups based on the above transformation behavior,
there are 16 which are tadpole free and give rise to the massless hypercharge\footnote{Note that setups with this
hypercharge have an additional symmetry under $b \rightarrow b'$.}.
Only two of these do not give rise to any R-parity violating
couplings on the perturbative level. Tables \ref{spectrum three stack model 1.1} and
\ref{spectrum three stack model 1.2} display for these two
setups the origin and the transformation behavior of the MSSM matter content.
 We analyze each case individually with respect to their Yukawa couplings\footnote{In \cite{Leontaris:2009ci} the author discusses a three-stack
 quiver similar to the ones we analyze here.}.
\begin{table}[h] \centering
\begin{tabular}{|c|c|c|c|c|}
\hline
 Sector & Matter fields &  Transformation & Multiplicity & Hypercharge\\
\hline \hline
 $ab$                            & $q_L$  & $(a,\overline{b})$ & $3 $& $\frac{1}{6}$ \\
\hline
 $ac'$                            & $u_R$  & $(\overline{a},\overline{c})$  & $3 $ & $-\frac{2}{3}$ \\
\hline
$aa'$ & $d_R$ & ${\Yasymm}_a$ & $3$& $\frac{1}{3}$  \\
\hline
$bc$                            & $L$  & $(b,\overline{c})$  & $3 $& $-\frac{1}{2}$ \\
\hline
$bc'$                            & $H_u+H_d$  & $(b,c)+(\overline{b},\overline{c})$ & $1 $ & $\frac{1}{2} \,\,\, -\frac{1}{2} $  \\
\hline
$bb'$                            & $N_R$  & $\overline{{\Yasymm}}_b $  & $3$ & $0$ \\
\hline
$cc'$                            & $E_R$  & ${\Ysymm}_c$  & $3 $ & $1$ \\
\hline
\end{tabular}
\caption{Spectrum for setup 1 with $U(1)_Y=\frac{1}{6}\,U(1)_a+\frac{1}{2}\,U(1)_c$} % \vspace{3mm}
\label{spectrum three stack model 1.1}
\end{table}\vspace{5pt}

Table \ref{spectrum three stack model
1.1} depicts the origin, transformation behavior, and
multiplicity of the respective matter fields for the first setup. One can easily check
that the constraints arising from tadpole cancellation and from the
presence of the massless hypercharge are satisfied. Moreover,
there are no R-parity violating couplings on the perturbative level.

The perturbatively allowed Yukawa couplings are
\begin{align*}
<{q^I_L}_{(1,-1,0)}\, {H_u}_{(0,1,1)} \, {u^J_R}_{(-1,0,-1)}>
\qquad
<L^I_{(0,1,-1)}\, {H_d}_{(0,-1,-1)} \, {E^J_R}_{(0,0,2)}>\\
\\
<L^I_{(0,1,-1)}\, {H_u}_{(0,1,1)} \, {N^J_R}_{(0,-2,0)}> \qquad
\qquad \,\,\,\,\, <{H_u}_{(0,1,1)}\,{H_d}_{(0,-1,-1)}>\,\,.
\end{align*}
Here the capital letters $I$ and $J$ denote the family index and the subscript
indicates the charge under the global $U(1)_a$, $U(1)_b$ and
$U(1)_c$. On the other hand the phenomenologically desired Yukawa
coupling
\begin{align}
<{q^I_L}_{(1,-1,0)}\, {H_d}_{(0,-1,-1)} \, {d^J_R}_{(2,0,0)}>
\end{align}
is perturbatively forbidden. An instanton carrying the charges
\begin{align}
Q_{a}(E_2)=-3 \qquad Q_{b}(E_2)=2 \qquad Q_c(E_2)=1
\end{align}
under the global $U(1)$'s can compensate for the overshooting in the
coupling $q_L\,H_d\,d_R$. Thus, in accord with \eqref{eq U(1)-charge of O(1) instanton}, a rigid $O(1)$ instanton exhibiting the intersection pattern
\begin{align}
I_{E2a}=1 \qquad I_{E2b}=-1 \qquad I_{E2c}=-1 \label{instanton d_r
example 1}
\end{align}
induces the coupling $q_L\,H_d\,d_R$. Let us analyze the path
integral in more detail. Apart from the generic uncharged zero modes
$x^{\mu}$ and $\theta^{\alpha}$, there are also three $ \ov
\lambda_a$, two $\lambda_b$ and one $\lambda_c$ zero modes.
\begin{figure}[h]
\begin{center}
 \includegraphics[width=0.7\textwidth]{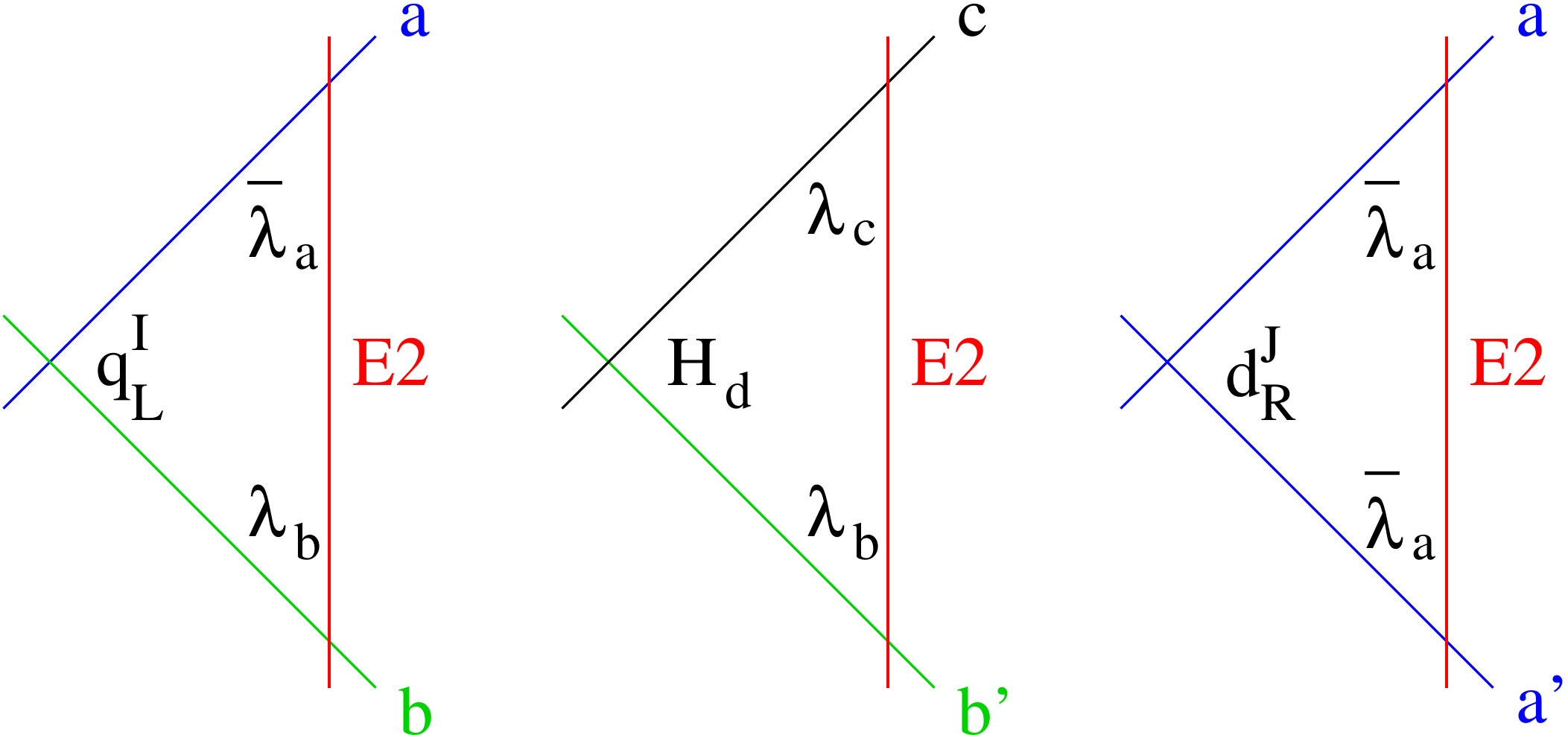}
\end{center}
\caption{\small Instanton induced Yukawa coupling $q^I_L\, H_d
\, d^J_R$ for setup 1.}\label{fig dquark}
\end{figure}
The path integral takes the form
\begin{align}
\int d^4 x \,d^2 \theta \, d^3 \ov \lambda_a \,d^2 \lambda_b \,d\lambda_c \,\,
e^{-S^{cl}_{E2}} <\ov \lambda_a q^I_{L} \lambda_b>\, <\lambda_b H_d
\lambda_c>\,<\ov \lambda_a d^J_R \ov \lambda_a> e^{Z'}\,\,,
\label{eq pathintegral}
\end{align}
where the three point amplitudes depicted in Figure \ref{fig dquark} can
be calculated applying CFT techniques \cite{Cvetic:2007ku}. Performing the
integral over the charged zero modes gives the superpotential
contribution
\begin{align}
\int d^4 x\, d^2 \theta \,Y^{IJ}_{q_L H_d d_R} \, q^I_L \, H_d \,d^J_R\,\,.
\end{align}
Here $Y^{IJ}_{q_L H_d d_R}$ contains the suppression factor
$e^{-S^{cl}_{E2}}$ of the instanton, the regularized one loop
amplitude $e^{Z'}$, as well as the world-sheet instanton contributions arising from the
three disc amplitudes. Note, though, that the induced $3\times 3$
Yukawa matrix factorizes
\begin{align}
Y^{IJ}_{q_L H_d d_R}=Y^{I}\, Y^{J}\,\,.
\end{align}
This is due to the fact that the disk amplitudes do not contain both
matter fields $q^I_L$ and $d^J_R$ simultaneously. Thus, in order to
generate non-vanishing masses for all three d-quarks, one needs
three different instantons with the intersection pattern
\eqref{instanton d_r example 1}.  Note that this not only explains the hierarchy
 between the u-quark Yukawa couplings, but also can account for the hierarchy
 between the d-quark families. To match experimental observations
the suppression factors should lie in the range of
$10^{-5}-10^{-2}.$\footnote{Note that the disk diagrams in
 \eqref{eq pathintegral} are suppressed by worldsheet instantons which may also
 contribute to the hierarchies.}

The Dirac neutrino masses are perturbatively realized and therefore are
expected to be of the same order as the masses for the other
leptons. The presence of a large Majorana mass for the right-handed
neutrinos would give a natural explanation for
the smallness of the neutrino masses via the seesaw mechanism. On the perturbative level such
a term is forbidden
\begin{align}
M_{N_R}\,{N_R}_{(0,-2,0)} \, {N_R}_{(0,-2,0)}\,\,.
\end{align}
A rigid $O(1)$-instanton with
the intersection pattern
\begin{align}
I_{E2a}=0 \qquad I_{E2b}=-2 \qquad I_{E2c}=0.
\end{align}
and a suppression factor of the instanton of order $10^{-5}$ to
$10^{-2}$ induces  such a Majorana mass term  in
the range $10^{13}\,GeV<M_{N_R}<10^{16}\,GeV$. Assuming that the Higgs VEV is around $100 \,GeV$,
one gets seesaw masses in the observed range $(10^{-3}-1)\,eV$.

Finally, let us mention that for this quiver another linear combination
\begin{align}
U(1)^{B-L}=-\frac{1}{6} \, U(1)_a - \frac{1}{2}\, U(1)_b+ \frac{1}{2} \, U(1)_c
\end{align}
satisfies the constraints \eqref{eq massless constraint non-abelian} and \eqref{eq massless constraint abelian} and might survive as local symmetry in the
low energy effective action. This linear combination can be interpreted as $U(1)^{B-L}$  and if indeed present would forbid  the generation of the Majorana mass term. In that case the  quiver is unrealistic, since their is no other mechanism to explain the smallness of the neutrino masses. Note though that the conditions \eqref{eq massless constraint non-abelian}
and \eqref{eq massless constraint abelian} are just necessary constraints for the presence of a massless U(1).
Whether such a linear combination is indeed massless depends on the concrete realization.
\vspace{0.5cm}

\begin{figure}[h]
\begin{center}
 \includegraphics[width=0.9\textwidth]{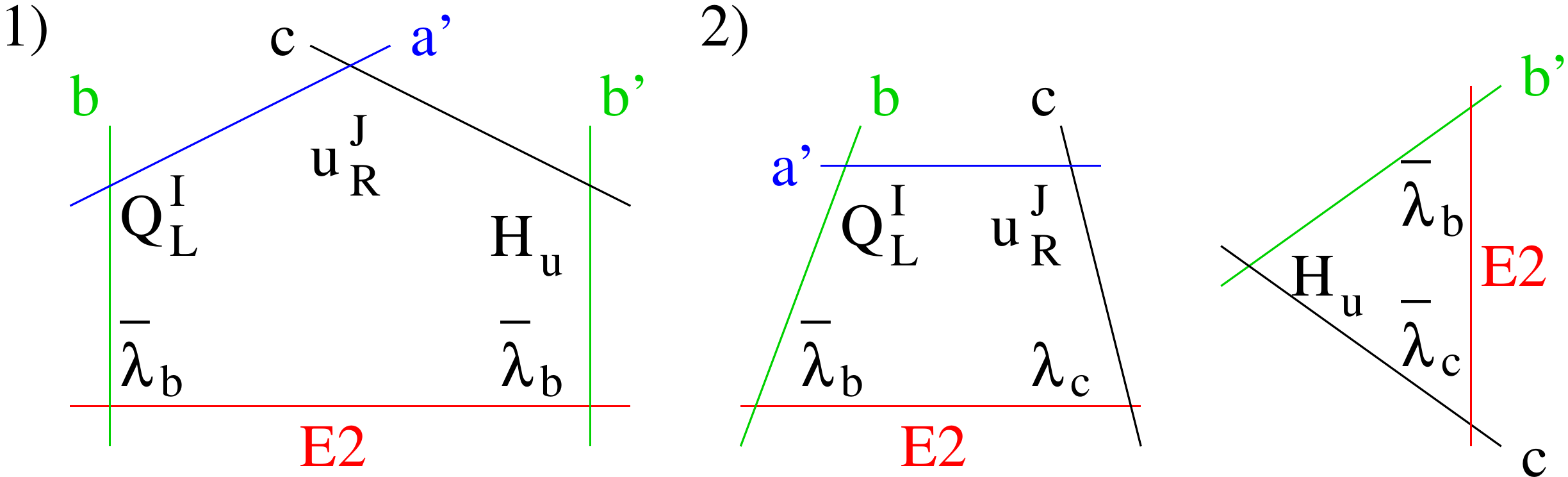}
\end{center}
\caption{\small Instanton induced Yukawa coupling $Q^I_L\, H_u
\, u^J_R$ for setup 2.}\label{fig upquark}
\end{figure}

We now turn to setup 2. Table \ref{spectrum three stack model 1.2}
displays the origin and transformation behavior of the MSSM matter
content.
\begin{table}[h] \centering
\begin{tabular}{|c|c|c|c|c|}
\hline
 Sector & Matter fields &  Transformation & Multiplicity & Hypercharge\\
\hline \hline
 $ab$                            & $q_L$  & $(a,\overline{b})$ & $1 $& $\frac{1}{6}$ \\
\hline
$ab'$                            & $Q_L$  & $(a,b)$ & $2 $& $\frac{1}{6}$ \\
\hline
 $ac'$                            & $u_R$  & $(\overline{a},\overline{c})$  & $3 $ & $-\frac{2}{3}$ \\
\hline
$aa'$ & $d_R$ & ${\Yasymm}_a$ & $3$& $\frac{1}{3}$  \\
\hline
$bc$                            & $L$  & $(b,\overline{c})$  & $3 $& $-\frac{1}{2}$ \\
\hline
$bc'$                            & $H_u+H_d$  & $(b,c)+(\overline{b},\overline{c})$ & $1 $ & $\frac{1}{2} \,\,\, -\frac{1}{2} $  \\
\hline
$bb'$                            & $N_R$  & ${\Yasymm}_b $  & $3$ & $0$ \\
\hline
$cc'$                            & $E_R$  & ${\Ysymm}_c$  & $3 $ & $1$ \\
\hline
\end{tabular}
\caption{Spectrum for setup 2 with $U(1)_Y=\frac{1}{6}\,U(1)_a+\frac{1}{2}\,U(1)_c$} % \vspace{3mm}
\label{spectrum three stack model 1.2}
\end{table}\vspace{5pt}
Here the perturbatively allowed couplings are
\begin{align*}
&<{q_L}_{(1,-1,0)}\, {H_u}_{(0,1,1)} \, {u^I_R}_{(-1,0,-1)}> \qquad
<L^I_{(0,1,-1)}\, {H_d}_{(0,-1,-1)} \, {E^J_R}_{(0,0,2)}>\\\\& \qquad\qquad\qquad\qquad\qquad
<{H_u}_{(0,1,1)} \, {H_d}_{(0,-1,-1)}>\,\,.
\end{align*}
The perturbatively forbidden, but phenomenologically desired couplings are
\begin{align*}
<{Q^I_{L}}_{(1,1,0)} \, {H_u}_{(0,1,1)} \, {u^J_R}_{(-1,0,-1)}>
\qquad <{Q^I_L}_{(1,1,0)}\, {H_d}_{(0,-1,-1)} \,
{d^J_R}_{(2,0,0)}>\\\\
<{q_L}_{(1,-1,0)}\, {H_d}_{(0,-1,-1)} \, {d^I_R}_{(2,0,0)}>\qquad
<L^I_{(0,1,-1)}\, {H_u}_{(0,1,1)} \, {N^J_R}_{(0,2,0)}>\,\,.
\end{align*}
Let us first discuss the u-quark Yukawa couplings. Before including
non-perturbative effects, the u-quark Yukawa coupling matrix takes
the form
\begin{align}
Y_{q_L H_u u_R}=\left(
\begin{array}{ccc}
A_{11} &  A_{12} &  A_{13} \\
0 & 0& 0 \\
0 & 0 & 0
\end{array}
\right) \label{up yukawa matrix}
\end{align}
and thus only one family acquires a mass. We
identify this family with the heaviest generation. In order to generate
masses for the other two families we have to fill the zero entries
in the Yukawa coupling matrix \eqref{up yukawa matrix}. There are
potentially two different instantons which could generate the missing coupling
$Q_L H_u u_R$. Their intersection
pattern is given by
\begin{align*}
I_{E2_1}&=0 \qquad I_{E2_1b}=1 \qquad I_{E2_1c}=0\\
I_{E2'_1a}&=0 \qquad I_{E2'_1b}=1 \qquad I_{E2'_1c}=0 \qquad
I^{{\cal N}=2}_{E2'_1c}=1\,\,.
\end{align*}
While the first instanton $E2_1$ exhibits only two charged zero
modes, namely two $\lambda_b$ modes, the other one $E2'_1$ has two
additional charged zero modes $\lambda_c$ and $\ov \lambda_c$.
Figure \ref{fig upquark} displays how these charged zero
modes are saturated via one or two disc diagrams, respectively.
For both types of instantons the induced Yukawa matrix does not
factorize, so one instanton can give masses to both families.
If both types of instantons are present, the one which wraps the smaller three-cycle in the
internal manifold and thus exhibits the smaller suppression
factor gives the dominant contribution.

\begin{figure}[h]
\begin{center}
 \includegraphics[width=0.7\textwidth]{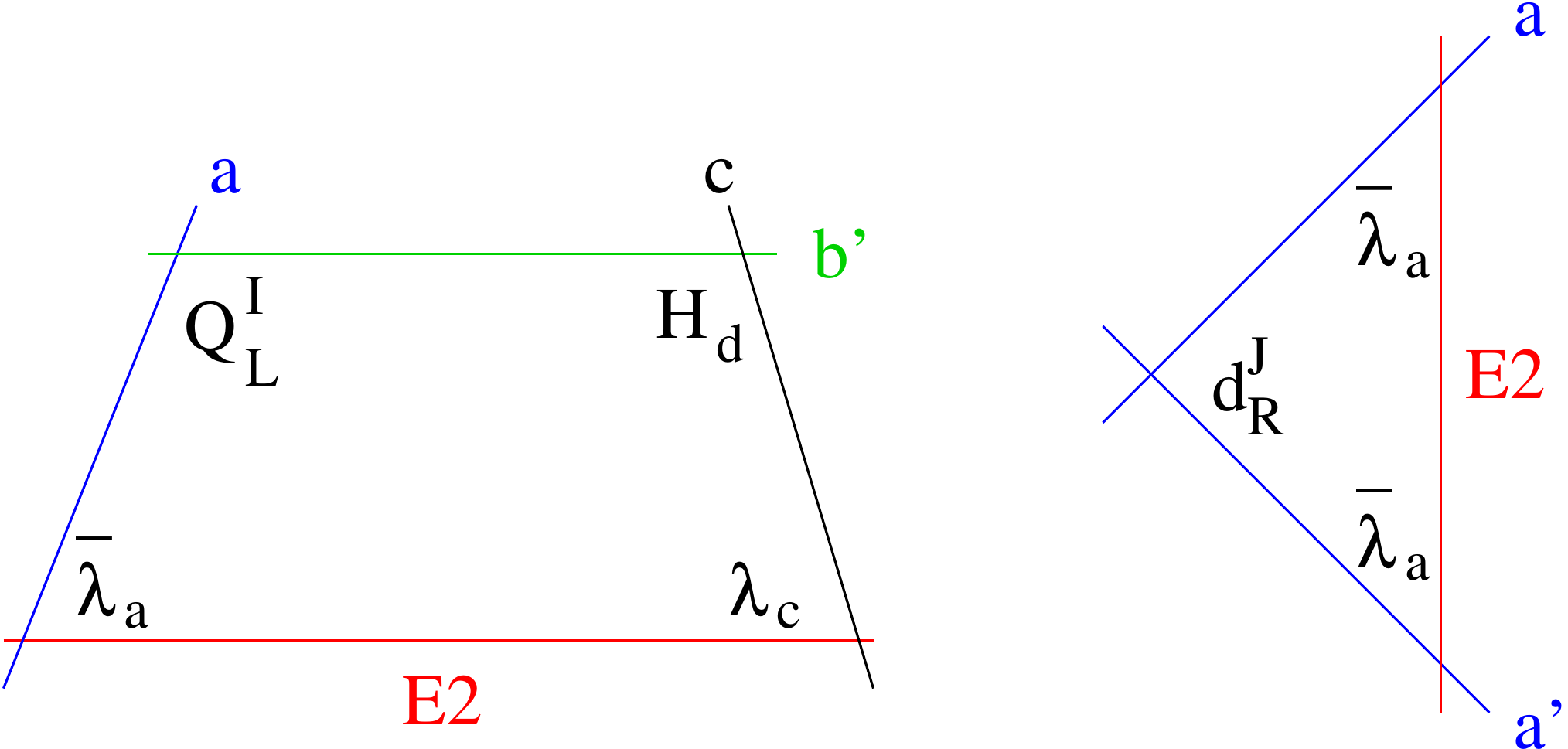}
\end{center}
\caption{\small Instanton induced Yukawa coupling $Q^I_L\, H_d
\, d^J_R$ for setup 2.}\label{fig dquark2}
\end{figure}

We now turn to the d-quark Yukawa coupling. Similarly to the
previous setup, the coupling $q_L\,H_d\,d^I_R$ gets generated by an
instanton with the intersection pattern
\begin{align*}
I_{E2_2a}=1 \qquad I_{E2_2b}=-1 \qquad I_{E2_2c}=-1.
\end{align*}
The coupling $Q^I_L\, H_d\,d^J_R$ is induced by an instanton with
the intersection pattern
\begin{align}
I_{E2_3a}=1 \qquad I_{E2_3b}=0 \qquad I_{E2_3c}=-1.
\label{intersectionpattern 2.1.1.2 }
\end{align}
The disk diagrams necessary to saturate the charged zero modes are
depicted in the Figures \ref{fig dquark} and \ref{fig dquark2}. Analogously to the previous setup,
the fact that the Yukawa matrix is factorizable requires at least two
different instantons with the intersection pattern
\eqref{intersectionpattern 2.1.1.2 }.

Looking at the quark Yukawa matrices after taking into account
perturbative and non-perturbative contributions, we see that all of quark masses
except the top mass are generated by instantons and are therefore suppressed
according to the volume of the instanton wrapped cycles.
Thus this quiver gives a natural explanation for the observed mass hierarchy of the top
quark relative to all the other quarks.

Finally, let us discuss the Dirac mass term $L^I H_u N^J_R$, which
can be generated via two types of instantons with intersection patterns
\begin{align}
&I_{E2_4a}=0 \qquad I_{E2_4b}=2 \qquad I_{E2_4c}=0\\
&I_{E2'_4a}=0 \qquad I_{E2'_4b}=2 \qquad I_{E2'_4c}=0\qquad I^{{\cal
N}=2}_{E2'_4c}=1.
\label{intersectionpattern 2.1.1.4}
\end{align}
As discussed in \cite{Cvetic:2008hi}, the Dirac neutrino
mass term is often perturbatively forbidden, but can be
generated by non-perturbative effects. Such a term would be suppressed and
would therefore give an intriguing explanation for the
smallness of the neutrino mass. For this quiver, however, the nonperturbative generation of Dirac mass term
via the instanton $E2_4$ cannot account for smallness of the neutrino mass since
$E2_4$ also generates a Majorana mass term. Taking both terms into account,
the mass matrix takes the form (here simplified for only one family)
\begin{align}
m_{\nu}= \left( \begin{array}{cc} 0 & e^{-S_{E2_4}} \langle H_u \rangle\\
e^{-S_{E2_4}}\langle H_u \rangle & e^{-S_{E2_4}} M_s \\
\end{array}\right)\,\,,
\end{align}
where $\langle H_u \rangle$ denotes the VEV of the Higgs field and $M_s$ is the
string mass. The mass eigenvalues of $m_{\nu}$ are of the order
\begin{align}
m^1_{\nu}= e^{-S_{E2_4}}\,\frac{{\langle H_u \rangle}^2}{M_s} \qquad \text{and}
\qquad m^2_{\nu}= e^{-S_{E2_4}}\,M_s
\end{align}
Taking $M_s \simeq 10^{18} \,GeV$, $\langle H_u \rangle \simeq 100\, GeV$, and a negligible
instanton suppression factor, we get
neutrino masses of order $10^{-5}\, eV$, which is lower than the experimental value.
Thus we need a suppression factor which is bigger
than 1, which is of course impossible\footnote{As for the Weinberg operator in principle
one can lower the string scale if one allows for large extra dimensions.}. This is very similar to
non-perturbative generation of the Weinberg operator \eqref{eq weinberg operator},
where the generic values for $\langle H_u \rangle$ and $M_s$ give a contribution
to the neutrino masses which is significantly smaller than the observed one.

On the other hand the instanton $E2'_4$, with the intersection
pattern $\eqref{intersectionpattern 2.1.1.4}$, does not generate a
Majorana mass term for the right-handed neutrinos. Thus if only $E2'_4$ is present and it
has a suppression factor of $10^{-14}$ to $10^{-11}$, then the non-perturbative
generation of the Dirac neutrino mass term indeed gives a natural
explanation for the smallness of the neutrino masses.

Note, though, that the instantons $E2_1$, $E2'_1$ and $E2_3$
generically induce the R-parity violating couplings $L\,L\,E_R$, $L
H_u$, $u_R\, d_R\, d_R$ and $q_L\,L\, d_R$ of the same order
as the induced Yukawa couplings $Q_{L} \, H_u \,u_R$ and $Q_{L} \, H_d\,
d_R$. Thus this setup is ruled out as unrealistic.

\subsubsection{$U(1)_Y=-\frac{1}{3} U(1)_a - \frac{1}{2} U(1)_b$
\label{sec threestack2}}

Now we turn to the other potential hypercharge choice for three-stack
models. For this hypercharge, the right-handed electron arises at
intersections between the brane $b$ and its image $b'$ and
transforms as antisymmetric of $SU(2)$. The right-handed neutrino is
located at intersections between $c$ and $c'$ and transforms as
symmetric under $U(1)_c$. The right-handed u-quarks transform as
antisymmetric of $SU(3)$, and thus appear at intersection of stack
$a$ with its orientifold image $a'$. The right-handed d-quarks have
potentially two origins, since they can arise from the sector $ac$ or
the sector $ac'$. In contrast to the previous case, the left-handed
quarks $q_L$ have only one possible origin, as they must arise from the
sector $ab$. Below we summarized the possible transformation
behaviors of the respective matter fields.
\begin{align*}
q_L\,:&\qquad (a,\ov{b})\\
u_R\,:&\qquad {\Yasymm}_a\\
d_R\,:&\qquad (\ov{a},c), \,\,\,\,\, (\ov{a},\ov{c})\\
L\,\,:&\qquad (b,\ov{c}), \,\,\,\,\, (b,c)\\
E_R\,:&\qquad \ov{\Yasymm}_b\\
N_R\,:&\qquad {\Ysymm}_c,\,\,\,\,\, \ov{\Ysymm}_c\\
H_u\,:&\qquad (\ov{b},c),\,\,\,\,\,(\ov{b},\ov{c})\\
H_d\,:&\qquad (b,\ov{c}),\,\,\,\,\,(b,c)
\end{align*}
If we require tadpole cancellation, masslessness of the hypercharge, and the absence of any R-parity violating
couplings, then we get again two different setups. We discuss each individually.

 \vspace{0.5cm}

For the first setup, the origin and transformation behavior of the
matter fields is given in Table \ref{spectrum three stack model
2.1}.
\begin{table}[h] \centering
\begin{tabular}{|c|c|c|c|c|}
\hline
 Sector & Matter fields &  Transformation & Multiplicity & Hypercharge\\
\hline \hline
 $ab$                            & $q_L$  & $(a,\overline{b})$ & $3$& $\frac{1}{6}$ \\
\hline
 $ac'$                            & $d_R$  & $(\overline{a},\overline{c})$  & $3 $ & $\frac{1}{3}$ \\
\hline
$aa'$ & $u_R$ & ${\Yasymm}_a$ & $3$& $-\frac{2}{3}$  \\
\hline
$bc$                            & $L$  & $( b,{\ov c})$  & $3 $& $-\frac{1}{2}$ \\
\hline
$bc'$                            & $H_u+H_d$  & $(\overline{b},\overline{c})+ (b,c)$ & $1 $ & $\frac{1}{2} \,\,\, -\frac{1}{2} $  \\
\hline
$bb'$                            & $E_R$  & $\overline{{\Yasymm}}_b $  & $3$ & $0$ \\
\hline
$cc'$                            & $N_R$  & ${\Ysymm}_c$    & $3$ & $1$ \\
\hline
\end{tabular}
\caption{Spectrum for setup 1 with $U(1)_Y=-\frac{1}{3} U(1)_a -
\frac{1}{2} U(1)_b$} % \vspace{3mm}
\label{spectrum three stack model 2.1}
\end{table}\vspace{5pt}
\noindent

Here, the perturbatively allowed couplings are
\begin{align*}
<{q^I_L}_{(1,-1,0)}\, {H_d}_{(0,1,1)} \, {d^J_R}_{(-1,0,-1)}>  \qquad
<L^I_{(0,1,-1)}\, {H_d}_{(0,1,1)} \, {E^J_R}_{(0,-2,0)}>\\
\\
<L^I_{(0,1,-1)}\, {H_u}_{(0,-1,-1)} \, {N^J_R}_{(0,0,2)}> \,\,\,\,\, \qquad
\qquad <{H_d}_{(0,1,1)}\,{H_u}_{(0,-1,-1)}>
\end{align*}
and the u-quark Yukawa coupling
\begin{align}
<{q^I_L}_{(1,-1,0)}\, {H_u}_{(0,-1,-1)} \, {u^J_R}_{(2,0,0)}>
\end{align}
is violating the global $U(1)$ symmetries and is therefore perturbatively
forbidden. It can be induced by an instanton with the intersection
pattern
\begin{align}
I_{E2_1a}=1 \qquad I_{E2_1b}=-1 \qquad I_{E2_1c}=-1.
\end{align}
Due to the factorization of the instanton induced Yukawa matrix, one
again needs three different instantons to generate mass terms for
all three generations.

The Dirac neutrino mass is perturbatively realized and expected to
be of the same order as the lepton masses. A large Majorana mass
\begin{align*}
M_{N_R}\,{N_R}_{(0,0,2)}\,{N_R}_{(0,0,2)}
\end{align*}
could account for the smallness of the neutrino masses via the seesaw mechanism.
Such a mass term can be generated by an instanton with the intersection pattern
\begin{align*}
I_{E2_2a}=0 \qquad I_{E2_2b}=0 \qquad I_{E2_2c}=4.
\end{align*}
In order to get neutrino masses compatible with experiments, the suppression factor should be in the range $10^{-5}$ to $10^{-2}$.

Both instantons, $E2_1$ and $E2_2$, do not generate any of the
dangerous R-parity violating couplings $d_R\, d_R\, u_R$,
$L\,L\,E_R$, $q_L \,L\, d_R$ or $L\,H_u$. Note,
though, that the u-quark coupling is realized non-perturbatively
while the d-quark couplings is perturbatively allowed. This suggests
that the u-quark masses are significantly smaller than
the d-quark masses, which contradicts experimental
observations.

This quiver may possess another massless $U(1)$ which is given by
\begin{align}
U(1)^{B-L}=-\frac{1}{6} \, U(1)_a - \frac{1}{2}\, U(1)_b+ \frac{1}{2} \, U(1)_c\,\,
\end{align}
and can be interpreted as $U(1)^{B-L}$. If this symmetry is present then the Majarana mass term cannot be generated and the quiver does not exhibit a mechanism which could account for the small
neutrino masses. Analogously to setup 1 in \ref{sec threestack1 }, the presence of such a massless linear combination depends crucially on the concrete realization.

\begin{table}[h] \centering
\begin{tabular}{|c|c|c|c|c|}
\hline
 Sector & Matter Fields &  Transformation & Multiplicity & Hypercharge\\
\hline \hline
 $ab$                            & $q_L$  & $(a,\overline{b})$ & $3$& $\frac{1}{6}$ \\
\hline
 $ac'$                            & $d_R$  & $(\overline{a},\overline{c})$  & $3 $ & $\frac{1}{3}$ \\
\hline
$aa'$ & $u_R$ & ${\Yasymm}_a$ & $3$& $-\frac{2}{3}$  \\
\hline
$bc$                            & $L$  & $( b,{\ov c})$  & $3 $& $-\frac{1}{2}$ \\
\hline
$bc'$                            & $H_u+H_d$  & $(\overline{b},\overline{c})+ (b,c)$ & $1 $ & $\frac{1}{2} \,\,\, -\frac{1}{2} $  \\
\hline
$bb'$                            & $E_R$  & $\overline{{\Yasymm}}_b $  & $3$ & $0$ \\
\hline
$cc'$                            & $N_R$  & $\ov {\Ysymm}_c$    & $3$ & $1$ \\
\hline
\end{tabular}
\caption{Spectrum for setup 2 with $U(1)_Y=-\frac{1}{3} U(1)_a -
\frac{1}{2} U(1)_b$} % \vspace{3mm}
\label{spectrum three stack model 2.2}
\end{table}\vspace{5pt}
\vspace{0.5 cm}
The second setup with this hypercharge which does not give rise to any
R-parity violating couplings differs from setup 1 only in the
transformation behavior of the right-handed neutrino $N_R$. Table
\ref{spectrum three stack model 2.2} summarizes the origin as well
as the transformation behavior of the MSSM matter content. The
perturbatively allowed couplings are
\begin{align*}
&<{q^I_L}_{(1,-1,0)}\, {H_d}_{(0,1,1)} \, {d^J_R}_{(-1,0,-1)}>  \qquad
<L^I_{(0,1,-1)}\, {H_d}_{(0,1,1)} \, {E^J_R}_{(0,-2,0)}>\\
\\
& \qquad \qquad \qquad \qquad \qquad<{H_u}_{(0,-1,-1)}\,{H_d}_{(0,1,1)}>\,\,,
\end{align*}
while here the Dirac neutrino mass term is perturbatively forbidden,
due to the different transformation behavior of the right-handed neutrino.
As in the first setup, the u-quark Yukawa coupling is perturbatively
absent. The perturbatively forbidden, but desired
couplings are
\begin{align}
<{q^I_L}_{(1,-1,0)}\, {H_u}_{(0,-1,-1)} \, {u^J_R}_{(2,0,0)}> \qquad
<L^I_{(0,1,-1)}\, {H_u}_{(0,-1,-1)} \, {N^J_R}_{(0,0,-2)}>\,\,.
\end{align}
The discussion of the non-perturbative generation of the u-quark
Yukawa coupling is analogous to setup 1. The coupling
$q_L\,H_u\,u_R$ can be induced by an instanton with the intersection
pattern
\begin{align}
I_{E2a}=1 \qquad I_{E2b}=-1 \qquad I_{E2c}=-1\,\,,
\end{align}
and one needs
three different instantons with such an intersection pattern
in order to generate masses for all three families. Again,
note that the non-perturbative generation of the top Yukawa coupling
is not favorable and a huge fine-tuning is required to match
experimental observations.

Let us turn to the Dirac neutrino mass term $L \,H_u\, N_R$. It can
be induced by an instanton with the intersection pattern
\begin{align}
I_{E2_2a}=0 \qquad I_{E2_2b}=0 \qquad I_{E2_2c}=-4\,\,.
\end{align}
Note, though, that the same instanton also induces a Majorana
mass term
\begin{align}
M_{N_R}\, {N_R}_{(0,0,-2)}\, {N_R}_{(0,0,-2)}
\end{align}
for the right handed neutrino, an effect similar to one seen in the second setup in
section \ref{sec threestack1 }.
In contrast to the quiver displayed in Table \ref{spectrum three stack model 1.2}, this setup does not exhibit an instanton which
only generates the Dirac mass term. Thus in this quiver we cannot account for the smallness
of the neutrino masses.

Let us summarize the phenomenological drawbacks of this setup once more. Though the
instantons required to generate the desired, but perturbatively
forbidden, couplings do not induce any R-parity violating couplings,
this setup still has two major flaws. First, the u-quark couplings are
generated non-perturbatively and are thus suppressed relative to
the d-quark couplings, which are perturbatively allowed.
Second, the instanton inducing the Dirac mass term for the
neutrinos also generates a Majorana mass term for the right-handed
neutrinos. Thus, the seesaw mechanism gives neutrino masses which are much smaller than
the observed values.

\subsection{Four-stack Models \label{sec four-stack models}}
Another very natural way of realizing the MSSM is to embed the
matter content at intersections of four stacks of D-branes, which wrap generic cycles and give rise
to the gauge symmetry
\begin{align}
U(3)_a\times U(2)_b \times U(1)_c \times U(1)_d\,\,.
\end{align}
The left-handed quarks $q_{L}$ are localized at the intersection of
brane $a$ with brane $b$ or its orientifold image $b'$. The
right-handed quarks, $u_R$ and $d_R$, arise at intersections of
brane $a$ with one of the $U(1)$ branes or its orientifold image, or at
the intersection of stack $a$ with its orientifold image, in which case
the right-handed quarks would transform as antisymmetrics of $SU(3)$. The left-handed leptons are charged
under the $U(2)$ and are neutral under $U(3)$, and thus appear at
intersections of brane $b$ with one of the $U(1)$ branes or its orientifold image.
Finally, the right-handed electron $E_R$ and the right-handed
neutrino $N_R$, both singlets under $U(3)$ and $U(2)$, arise at
the intersection of two $U(1)$ branes or at the intersection of $b$ with $b'$, in which case they would
transform as antisymmetric of $SU(2)$.

For the four-stack models, the hypercharges compatible with the MSSM hypercharge
assignment are \cite{Anastasopoulos:2006da}
\begin{itemize}
\item[$\bullet$] $U(1)_Y=-\frac{1}{3} U(1)_a - \frac{1}{2} U(1)_b  $
\item[$\bullet$] $U(1)_Y= -\frac{1}{3} U(1)_a -\frac{1}{2} U(1)_b - \frac{1}{2} U(1)_d $
\item[$\bullet$] $U(1)_Y= -\frac{1}{3} U(1)_a -\frac{1}{2} U(1)_b + U(1)_d $
\item[$\bullet$] $U(1)_Y=\frac{1}{6} U(1)_a +\frac{1}{2} U(1)_c$
\item[$\bullet$] $U(1)_Y= \frac{1}{6} U(1)_a +\frac{1}{2} U(1)_c -\frac{1}{2}U(1)_d$
\item[$\bullet$] $U(1)_Y= \frac{1}{6} U(1)_a +\frac{1}{2} U(1)_c - \frac{3}{2}U(1)_d$\,\,.
\end{itemize}
Before discussing the different hypercharge setups in detail, let us
summarize what we call a phenomenologically favorable setup:
\begin{itemize}
\item[$\bullet$] All the MSSM matter content and the right-handed neutrinos, apart from the Higgs
fields, appear as chiral fields at an intersection between two
D-branes.
We emphasize again that all the MSSM matter and the right-handed
neutrinos have to appear at intersections between the four D-branes
and their orientifold images. Moreover, there are no further chiral
fields charged under the gauge groups of the four D-branes.
\item[$\bullet$] As discussed in chapter \ref{sec constraints}, the tadpole
condition puts restrictions on the transformation behavior of the
matter fields. We require these constraints to be satisfied.
\item[$\bullet$] In chapter \ref{sec constraints}, we derived
constraints on the transformation properties of the matter fields arising from the presence
of a massless $U(1)$ gauge symmetry. We require that these
constraints, given by equations
\eqref{eq massless constraint non-abelian} and
\eqref{eq massless constraint abelian}, are satisfied by the hypercharge $U(1)_Y$.
\item[$\bullet$] We forbid any R-parity violating couplings on
perturbative level. These include the couplings $L\, L\,
E_R$, $u_R\, d_R\,d_R$, $q_L\, L \, d_R$ and $ L\, H_u $.
\item[$\bullet$]
All the Yukawa couplings which are missing, due to being perturbatively forbidden,
are generated by instantons. The desired couplings are:
\begin{align*}
q_L\, H_u\,u_R \qquad q_L\, H_d\,d_R\qquad L\, H_d\,E_R \qquad L\,
H_u\,N_R \qquad H_u\, H_d \,\,.
\end{align*}
We require that all three families of the u-quark, d-quark, and
electron acquire a mass. This translates into the condition that the associated
Yukawa matrices, $Y^{IJ}_{q_L\, H_u\,u_R}$, $Y^{IJ}_{q_L\, H_d\,d_R}$
or $Y^{IJ}_{L\, H_d\,E_R}$ have non-zero eigenvalues, where the entries of the respective Yukawa matrix can be perturbatively or
non-perturbatively generated. For the neutrinos we allow one generation to be light, or even
massless. Thus the Yukawa matrix $Y^{IJ}_{L\, H_u\,N_R }$may
exhibit one zero eigenvalue. The other two eigenvalues must be
non-vanishing.
\item[$\bullet$] Often times an instanton which is required to generate the Yukawa couplings
also induces a tadpole $N_R$ and thus an instability for the setup. We rule out any setup which requires the presence of such an instanton.
\item[$\bullet$]
As seen in \ref{sec threestack1 }, an instanton which is required to generate
the Yukawa couplings might also generate an R-parity violating coupling of the same order. Quivers which require such instantons are ruled out as unrealistic.
\item[$\bullet$]
We rule out setups which lead to a large family mixing in the quark Yukawa couplings.
For example, this might
happen if the left-handed quark has two possible origins, namely the
$ab$ sector and the $ab'$ sector. As encountered in \cite{Ibanez:2008my,Leontaris:2009ci} the quark Yukawa matrices, taking into account
only perturbative contributions, take the form
\begin{align}
Y^P_{u^I_L H_u u^J_R}=\left(\begin{array}{ccc} A^u_{11}&A^u_{12}&
A^u_{13}\\
A^u_{21}&A^u_{22}& A^u_{23}\\
0&0&0\\
\end{array}\right)
\qquad Y^{P}_{d^I_L H_d d^J_R}=\left(\begin{array}{ccc}
0&0&0\\
0&0&0\\
A^d_{31}&A^d_{32}& A^d_{33}\\
\end{array}\right)\,\,.
\end{align}
Such a Yukawa texture suggests that the u-quark mass for the two
heaviest families is much larger than the one for the lightest
family after also taking into account instanton effects which will
fill the zero entries in the matrix $Y^P_{u^I_L H_u u^J_R}$. On the
other hand it also suggests that the family with the lightest u-quark also contains
the heaviest d-quark, in contradiction with experimental observation.
Setups with such Yukawa matrix texture\footnote{Analogously we rule out other setups which exhibit Yukawa textures that also lead to similar problems, for instance the following Yukawa textures:
\begin{align*}
Y^P_{u^I_L H_u u^J_R}=\left(\begin{array}{ccc} A^u_{11}&0&
0\\
0&0&0\\
0&0&0\\
\end{array}\right)
\qquad Y^{P}_{d^I_L H_d d^J_R}=\left(\begin{array}{ccc}
0&0&0\\
A^d_{21}&0&0\\
A^d_{31}&0&0\\
\end{array}\right)\,\,.
\end{align*} lead to a large, undesired family mixing.}
lead to a large mixing between different families
which is not observed in nature. Therefore we rule out such setups,
since they are phenomenologically unrealistic.
\item[$\bullet$] If the $\mu$-term is perturbatively forbidden,
we require it to be generated non-perturbatively. If an instanton simultaneously generates a perturbatively forbidden, but desired, Yukawa coupling
and the $\mu$ term, the latter is generically too large. Such a situation was encountered in
 \cite{Ibanez:2008my}, where it was shown that this problem can be overcome
if one allows for a second Higgs pair. In this work we
do not allow for a second Higgs pair, since we restrain ourselves to
the exact MSSM matter content plus three right-handed neutrinos.
We leave it for future work to make a systematic analysis
where one allows for a second Higgs pair.
\item[$\bullet$] We require that a phenomenologically realistic setup
exhibits a natural explanation for the smallness of the neutrino
mass. As discussed in section \ref{sec neutrino masses} and in
\cite{Ibanez:2007rs}, the non-perturbative realization
of the Weinberg operator does not give neutrino masses in the
desired range unless the string mass is significantly lowered.

In \cite{Cvetic:2008hi}, the authors present an intriguing
mechanism which is based on the non-perturbative generation of the
Dirac mass term for the neutrinos.
We saw a realization of this mechanism in the second setup of section \ref{sec threestack1 }. For
this mechanism to work, instantons inducing a Majorana mass term for
the right-handed neutrinos must be absent. In particular, the
same instanton which generates the Dirac mass term must not also
generate the Majorana mass term. Otherwise the neutrino masses are not in the observed
range,  as we saw in section
\ref{sec threestack2}.

Another possibility to obtain small neutrino masses is via the
seesaw mechanism. A necessary ingredient for this mechanism is a
large Majorana mass term for the right-handed neutrinos, which can
be generated non-perturbatively \cite{Blumenhagen:2006xt,Ibanez:2006da,Cvetic:2007ku,Ibanez:2007rs,Antusch:2007jd,Cvetic:2007qj}.
We impose the same constraints on the Majorana mass generating instantons that
we impose on instantons which generate Yukawa couplings.
Specifically, we require these instantons do not generate R-parity violating
couplings, $N_R$ tadpoles, or give rise to a $\mu$-term that is too large.
\item[$\bullet$] Due to the fact that the top quark mass is so much
larger than all other masses of the elementary particles, we require
that its Yukawa coupling is realized perturbatively. The
non-perturbative suppression of other couplings relative to the top Yukawa coupling
then gives a natural explanation for the
observed hierarchies, without too much fine-tuning. We also allow for quivers
where none of the Yukawa couplings $q_L\,H_u\, u_R$,
$q_L\,H_d\, d_R$ and $L \, H_d\, E_R$ is perturbatively realized. For such quivers
one can still obtain a natural hierarchy between the top quark mass and all other
matter fields if the instanton inducing the top Yukawa coupling is least suppressed.
\end{itemize}
For every choice of hypercharge, we present the potential transformation behavior under the gauge groups
and give the number of setups which satisfy the various phenomenological conditions listed
above. The small subset of quivers which are compatible with all these constraints are listed in tables.
The latter represent a good starting point for concrete string realizations of the MSSM with realistic Yukawa textures.

\subsubsection{$U(1)_Y= -\frac{1}{3} U(1)_a -\frac{1}{2} U(1)_b - \frac{1}{2} U(1)_d $}

\begin{align*}
q_L\,:&\qquad (a,\ov{b})\\
u_R\,:&\qquad \Yasymm_a\\
d_R\,:&\qquad (\ov{a},c),\,\,\,\, (\ov{a},\ov c)\\
L\,\,:&\qquad (b,\ov{c}),\,\,\,\,(b,c)\\
E_R\,:&\qquad \ov {\Yasymm}_b,\,\,\,\,\ov\Ysymm_d\\
N_R\,:&\qquad {\Ysymm}_c,\,\,\,\,\ov{\Ysymm}_c\\
H_u\,:&\qquad (\ov{b},c),\,\,\,\,(\ov b,\ov{c})\\
H_d\,:&\qquad (b,\ov{c}),\,\,\,\,(b,c)\\
\end{align*}
There exist a few solutions that satisfy the tadpole and masslessness conditions. For all these solutions  none of the matter fields is charged under $U(1)_d$. Thus they correspond to the three-stack quivers analyzed in section \ref{sec threestack2}. For this and all following choices of hypercharge we require that all four D-branes are populated. Therefore for this choice of hypercharge we do not find any solutions which satisfy the tadpole and masslessness constraints.

\subsubsection{$U(1)_Y=\frac{1}{6} U(1)_a + \frac{1}{2} U(1)_c$}

\begin{align*}
q_L\,:&\qquad (a,\ov{b}),\,\,\,\,(a,b)\\
u_R\,:&\qquad (\ov{a},\ov{c})\\
d_R\,:&\qquad (\ov{a},c),\,\,\,\,{\Yasymm}_a\\
L\,\,:&\qquad (b,\ov{c}),\,\,\,\,(\ov{b},\ov{c})\\
E_R\,:&\qquad {\Ysymm}_c\\
N_R\,:&\qquad {\Yasymm}_b,\,\,\,\,\ov{\Yasymm}_b,\,\,\,\,{\Ysymm}_d,\,\,\,\,\ov{\Ysymm}_d\\
H_u\,:&\qquad (\ov{b},c),\,\,\,\,(b,c)\\
H_d\,:&\qquad (b,\ov{c}),\,\,\,\,(\ov{b},\ov{c})\\
\end{align*}

This hypercharge exhibits 8 tadpole free setups with the exact MSSM
matter content and a massless $U(1)_Y$. Out of these, only one does not give
rise to R-parity violating couplings on the perturbative level. If
one includes the instanton required to induce the desired Yukawa
couplings, the same instanton will also generate R-parity violating couplings,
and thus this setup is unrealistic.

\subsubsection{$U(1)_Y= \frac{1}{6} U(1)_a +\frac{1}{2} U(1)_c -
\frac{3}{2}U(1)_d $}
\begin{align*}
q_L\,:&\qquad (a,\ov{b}),\,\,\,\,(a,b)\\
u_R\,:&\qquad (\ov{a},\ov{c})\\
d_R\,:&\qquad (\ov{a},c),\,\,\,\,{\Yasymm}_a\\
L\,\,:&\qquad (b,\ov{c}),\,\,\,\,(\ov{b},\ov{c})\\
E_R\,:&\qquad {\Ysymm}_c,\,\,\,\,(\ov{c},\ov{d})\\
N_R\,:&\qquad {\Yasymm}_b,\,\,\,\,\ov{\Yasymm}_b\\
H_u\,:&\qquad (\ov{b},c),\,\,\,\,(b,c)\\
H_d\,:&\qquad (b,\ov{c}),\,\,\,\,(\ov b,\ov{c})\\
\end{align*}

For this hypercharge, one obtains 16 models with the exact MSSM matter content which also
satisfy the constraints arising from tadpole cancellation and the
masslessness of the hypercharge. Only 2 of them do not give rise to
any R-parity violating couplings on the perturbative level. Moreover,
one of these is unrealistic since the instanton
generating the desired, but perturbatively forbidden, Yukawa
couplings also induces R-parity violating couplings. The only
surviving setup is displayed in Table \ref{table spectrum four stack model
1/6a+1/2c-3/2d} \footnote{Let us briefly explain how to read the table.
Any given row specifies one solution, and the columns display the potential transformation
behaviors of the MSSM matter content giving rise to right charge under the Standard model
gauge groups.  For every solution the table indicates how many matter fields have the respective
transformation behavior. Note that the choice of  hypercharge may lead to some symmetries within the
MSSM spectrum. For instance  any solution for $U(1)_Y=\frac{1}{6}\,U(1)_a+\frac{1}{2}\,U(1)_c-\frac{3}{2} U(1)_d$
is also a solution under the exchange of  $b \rightarrow b'$. We take into account these types of symmetries
and only present nonequivalent solutions.} .

This quiver allows for an additional massless $U(1)$, satisfying \eqref{eq massless constraint non-abelian}
and \eqref{eq massless constraint abelian}, which is given by
\begin{align}
U(1)^{add}=\frac{2}{3} \, U(1)_a + U(1)_b\,\,.
\label{eq: additional massless U(1)}
\end{align}
Note though that the constraints \eqref{eq massless constraint non-abelian} and \eqref{eq massless constraint abelian}
are just necessary conditions, and whether $U(1)^{add}$ is indeed massless needs to be checked for a concrete realization.
Assuming $U(1)^{add}$ is massless the gauge symmetries in four dimensional spacetime is
\begin{align}
SU(3)\times SU(2) \times U(1)_Y \times U(1)^{add}\,\,,
\end{align}
which would forbid the generation of a Majorana mass term for the right-handed neutrinos. Since the Dirac neutrino mass is perturbatively realized,
$U(1)^{add}$ being massless implies that the seesaw does not work, and there is no mechanism to account for the smallness of the neutrino mass.

Subsequently, for the other choices of hypercharge, we will encounter for a few quivers a similar scenario,
where the presence of an additional massless $U(1)$ might forbid desired couplings.
We will denote such quivers with a $\dagger$. Let us emphasize again that even though the linear combination \eqref{eq: additional massless U(1)}
does satisfy the constraints \eqref{eq massless constraint non-abelian} and \eqref{eq massless constraint abelian}, this is not sufficient to conclude that $U(1)^{add}$ is indeed massless for a concrete realization.

\begin{table}[h]
\centering
\scalebox{.70}{
\begin{tabular}{|c|c|c|c|c|c|c|c|c|c|c|c|c|c|c|c|}\hline
		\multirow{2}{*}{Solution \#}&\multicolumn{2}{|c}{$q_L$} & \multicolumn{2}{|c|}{$d_R$} & $u_R$ & \multicolumn{2}{|c}{$L$} & \multicolumn{2}{|c}{$E_R$}
		& \multicolumn{2}{|c|}{$N_R$} & \multicolumn{2}{|c|}{$H_u$}  & \multicolumn{2}{|c|}{$H_d$} \\
        \cline{2-16}
		&$(a,b)$ & $(a,\ov{b})$ & $(\ov{a},c)$ & ${\Yasymm}_a$ & $(\ov{a},\ov{c})$ &$(b,\ov{c})$ &$(\ov{b},\ov{c}) $
		 & $(\ov{c},\ov{d})$ & ${\Ysymm}_c$ & ${\Yasymm}_b$ & $\ov{\Yasymm}_b$ & $(\ov b, c)$&$(b,c)$ & $(b,\ov{c})$
		& $(\ov{b},\ov{c}) $\\
		\hline\hline			
		1$^{\dagger}$&0&3&0&3&3&3&0&3&0&0&3&0&1&0&1\\ \hline		
\end{tabular}}
\caption{Spectrum for the solution with $U(1)_Y=\frac{1}{6}U(1)_a+\frac{1}{2}U(1)_c-\frac{3}{2} U(1)_d$.
\label{table spectrum four stack model
1/6a+1/2c-3/2d}}
\end{table}\vspace{5pt}

\subsubsection{$U(1)_Y= -\frac{1}{3} U(1)_a -\frac{1}{2} U(1)_b $}

\begin{align*}
q_L\,:&\qquad (a,\ov{b})\\
u_R\,:&\qquad {\Yasymm}_a\\
d_R\,:&\qquad (\ov{a},c),\,\,\,\,(\ov{a},\ov{c}),\,\,\,\,(\ov{a},d),\,\,\,\,(\ov{a},\ov{d})\\
L\,\,:&\qquad (b,\ov{c}),\,\,\,\,(b,c),\,\,\,\,(b,\ov{d}),\,\,\,\,(b,d)\\
E_R\,:&\qquad \ov{\Yasymm}_b\\
N_R\,:&\qquad (c,\ov{d}),\,\,\,\,(\ov{c},d),\,\,\,\,(c,d),\,\,\,\,(\ov{c},\ov{d}),\,\,\,\,
{\Ysymm}_c,\,\,\,\,\ov{\Ysymm}_c,\,\,\,\,{\Ysymm}_d,\,\,\,\,\ov{\Ysymm}_d\\
H_u\,:&\qquad (\ov{b},c),\,\,\,\,(\ov{b},\ov{c}),\,\,\,\,(\ov{b},d),\,\,\,\,(\ov{b},\ov{d})\\
H_d\,:&\qquad (b,\ov{c}),\,\,\,\,(b,c),\,\,\,\,(b,\ov{d}),\,\,\,\,(b,d)\\
\end{align*}

Allowing for an additional D-brane compared to the three-stack
setups with the same hypercharge, discussed in section \ref{sec threestack2}, gives rise to
782 models which satisfy the tadpole cancellation and hypercharge masslessness constraints. 135
of these setups do not exhibit any R-parity violating couplings
on the perturbative level. Only 44 models turn out to be realistic
once we take into account non-pertubative constraints on the instantons
which generate the desired Yukawa couplings.

Note that the u-quarks for this hypercharge appear as
antisymmetric of $SU(3)$, and therefore the top Yukawa coupling is never
realized perturbatively. However, there are 12 models in which
all the Yukawa couplings $q_L \, H_u u_R$, $q_L \, H_d \, d_R$ and
$L \, H_d \,E_r$ are perturbatively forbidden. For these setups it
is still possible to generate a hierarchy
between the top Yukawa coupling and all other couplings
without too much fine-tuning.

It turns out that 3 of these 12 models require a large fine-tuning to generate neutrino
masses in the desired range. This is due to the fact that the instantons
which induce the Dirac mass terms also generate  Majorana mass terms.
This was already encountered in section \ref{sec threestack2}, where
we have shown that in such a case the seesaw mechanism generates neutrino masses
which are too small. All twelve solutions are displayed in Table \ref{table four stack model  1/3a - 1/2 c}, where we indicate with a $^\bullet$
the 3 solutions which need a large fine-tuning to obtain neutrino masses in the desired range\footnote{In Table \ref{table four stack model  1/3a - 1/2 c}, we omit columns which have zero
entries for all solutions.}.

Four quivers, denoted by $\dagger$, may exhibit additional massless $U(1)$'s
which forbid the non-perturbative generation of perturbatively forbidden, but desired couplings.
In this case, such setups have to be ruled out as unrealistic.

\begin{table}[h]
\hspace{-.5cm}
\scalebox{.75}{
\begin{tabular}{|c|c|c|c|c|c|c|c|c|c|c|c|c|c|c|c|}
		\cline{1-15}
		\multirow{2}{*}{Solution \#}&\multicolumn{1}{|c}{$q_L$} & \multicolumn{1}{|c|}{$d_R$} & \multicolumn{1}{|c|}{$u_R$} & \multicolumn{1}{|c}{$L$} & \multicolumn{1}{|c}{$E_R$}
		& \multicolumn{7}{|c|}{$N_R$} & \multicolumn{1}{|c|}{$H_u$} & \multicolumn{1}{|c|}{$H_d$} \\
       \cline{2-15}
		& $(a,\ov{b})$ &$(\ov{a},\ov{d})$ &${\Yasymm}_a$& $(b,\ov{d})$
		& $\ov{\Yasymm}_b$ & $(c,d)$ & $(\ov{c},d)$ & $(\ov{c},\ov{d})$& ${\Ysymm}_c$ &$\ov{\Ysymm}_c$ &
		${\Ysymm}_d$ &$\ov{\Ysymm}_d$ & $(\ov{b},\ov{c})$ & $(b,c)$ \\
		\hline\hline		
1$^{\dagger}$&3&3&3&3&3&0&0&0&0&0&0&3&1&1\\ \hline
2$^{\dagger}$&3&3&3&3&3&0&0&0&0&0&3&0&1&1\\ \hline
3$^{\bullet\dagger}$&3&3&3&3&3&0&0&3&0&0&0&0&1&1\\ \hline
4&3&3&3&3&3&0&3&0&0&0&0&0&1&1\\ \hline
5$^{\dagger}$&3&3&3&3&3&1&1&0&0&0&0&1&1&1\\ \hline
6&3&3&3&3&3&3&0&0&0&0&0&0&1&1\\ \hline
7$^\bullet$&3&3&3&3&3&0&0&1&0&1&0&1&1&1\\ \hline
8&3&3&3&3&3&0&1&0&0&1&1&0&1&1\\ \hline
9&3&3&3&3&3&0&0&0&0&3&0&0&1&1\\ \hline
10$^\bullet$&3&3&3&3&3&0&1&1&1&0&0&0&1&1\\ \hline
11&3&3&3&3&3&1&0&0&1&0&1&0&1&1\\ \hline
12&3&3&3&3&3&0&0&0&3&0&0&0&1&1\\ \hline
\end{tabular}
}
\caption{Spectrum for the solution with $U(1)_Y= -\frac{1}{3} U(1)_a -\frac{1}{2} U(1)_b $.
\label{table four stack model  1/3a - 1/2 c}}
\end{table}

\subsubsection{$U(1)_Y= -\frac{1}{3}U(1)_a -\frac{1}{2} U(1)_b + U(1)_d $}

\begin{align*}
q_L\,:&\qquad (a,\ov{b})\\
u_R\,:&\qquad (\ov{a},\ov{d}),\,\,\,\,{\Yasymm}_a\\
d_R\,:&\qquad (\ov{a},c),\,\,\,\,(\ov{a},\ov{c})\\
L\,\,:&\qquad (b,\ov{c}),\,\,\,\,(b,c),\,\,\,\,(\ov{b},\ov{d})\\
E_R\,:&\qquad \ov{\Yasymm}_b,\,\,\,\,(\ov{c},d),\,\,\,\,(c,d)\\
N_R\,:&\qquad {\Ysymm}_c,\,\,\,\,\ov{\Ysymm}_c\\
H_u\,:&\qquad (\ov{b},c),\,\,\,\,(\ov{b},\ov{c}),\,\,\,\,(b,d)\\
H_d\,:&\qquad (b,\ov{c}),\,\,\,\,(b,c),\,\,\,\,(\ov{b},\ov{d})\\
\end{align*}

With this choice of hypercharge there are 144 tadpole free setups
with exactly the MSSM matter content and a massless $U(1)_Y$. Out of these, only 30 do not
exhibit any R-parity violating couplings on the perturbative level. Moreover, if we also take
into account non-perturbative effects, an additional 8 setups are ruled
out. Thus we obtain 22 solutions which not only are tadpole free and
do not generate any R-parity violating couplings, but also can
generate a $\mu$ term of the desired order. In order to explain the hierarchy between the top-quark mass and all other matter fields we further require the absence of the Yukawa couplings $q_L \, H_d \, d_R$ and
$L \, H_d \,E_r$ on the perturbative level. We obtain 14 models that are displayed in Table \ref{table four stack model  1/3a - 1/2 b+1 d}.
\begin{table}[h]
\centering
\scalebox{.8}{
\begin{tabular}{|c|c|c|c|c|c|c|c|c|c|c|c|c|} \hline
		\multirow{2}{*}{Solution \#}&\multicolumn{1}{|c}{$q_L$} & \multicolumn{1}{|c|}{$d_R$} & \multicolumn{2}{|c|}{$u_R$} & \multicolumn{1}{|c}{$L$} & \multicolumn{2}{|c}{$E_R$}
		& \multicolumn{2}{|c|}{$N_R$} & \multicolumn{1}{|c|}{$H_u$} & \multicolumn{2}{|c|}{$H_d$} \\
       \cline{2-13}
		& $(a,\ov{b})$ & $(\ov{a},\ov{c})$ & $(\ov{a},\ov{d})$ &${\Yasymm}_a$ & $(b,\ov{c})$
		& $(c,d)$ & $\ov{\Yasymm}_b$ & ${\Ysymm}_c$ & $\ov{\Ysymm}_c$ & $(b,d)$
		& $(b,c)$ &$(\ov{b},\ov{d})$ \\
		\hline\hline			
1&3&3&3&0&3&1&2&0&3&1&1&0\\ \hline
2&3&3&3&0&3&1&2&3&0&1&1&0\\ \hline
3&3&3&3&0&3&0&3&0&3&1&0&1\\ \hline
4&3&3&3&0&3&0&3&3&0&1&0&1\\ \hline
5&3&3&2&1&3&1&2&0&3&1&1&0\\ \hline
6&3&3&2&1&3&1&2&3&0&1&1&0\\ \hline
7&3&3&2&1&3&0&3&0&3&1&0&1\\ \hline
8&3&3&2&1&3&0&3&3&0&1&0&1\\ \hline
9&3&3&1&2&3&1&2&0&3&1&1&0\\ \hline
10&3&3&1&2&3&1&2&3&0&1&1&0\\ \hline
11&3&3&1&2&3&0&3&0&3&1&0&1\\ \hline
12&3&3&1&2&3&0&3&3&0&1&0&1\\ \hline
13$^{\dagger}$&3&3&0&3&3&0&3&0&3&1&0&1\\ \hline
14$^{\dagger}$&3&3&0&3&3&0&3&3&0&1&0&1\\ \hline
\end{tabular}}
\caption{Spectrum for the solutions with $U(1)_Y= -\frac{1}{3} U(1)_a -\frac{1}{2} U(1)_b + U(1)_d $.
\label{table four stack model  1/3a - 1/2 b+1 d}
}
\end{table}
The last two quivers in Table \ref{table four stack model  1/3a - 1/2 b+1 d} may give rise to an additional
massless $U(1)$ which if indeed present forbids some of the desired Yukawa couplings.

\subsubsection{$U(1)_Y= \frac{1}{6} U(1)_a +\frac{1}{2} U(1)_c -\frac{1}{2}U(1)_d $}

\begin{align*}
q_L\,:&\qquad (a,\ov{b}),\,\,\,\,(a,b)\\
u_R\,:&\qquad (\ov{a},\ov{c}),\,\,\,\,(\ov{a},d)\\
d_R\,:&\qquad {\Yasymm}_a,\,\,\,\,(\ov{a},c),\,\,\,\,(\ov{a},\ov{d})\\
L\,\,:&\qquad (b,\ov{c}),\,\,\,\,(\ov{b},\ov{c}),\,\,\,\,(\ov b,d),\,\,\,\,(b,d)\\
E_R\,:&\qquad (c,\ov{d}),\,\,\,\,{\Ysymm}_c,\,\,\,\,\ov{\Ysymm}_d\\
N_R\,:&\qquad {\Yasymm}_b,\,\,\,\,\ov{\Yasymm}_b\,,\,\,\,(c,d),\,\,\,\, (\ov c,\ov d)  \\
H_u\,:&\qquad (\ov b,c),\,\,\,\,(b,c),\,\,\,\,(b,\ov{d}),\,\,\,\,(\ov{b},\ov{d})\\
H_d\,:&\qquad (b,\ov{c}),\,\,\,\,(\ov{b},\ov{c}),\,\,\,\,(\ov b,d),\,\,\,\,(b,d)\\
\end{align*}

For this choice of massless hypercharge we obtain 3974 tadpole free
models, only 480 do not give rise to any R-parity violating couplings at the perturbative level.
Including non-perturbative effects the number, of realistic models is
51. Moreover, if we require that the top Yukawa coupling is
perturbatively present or, in case it is absent, that the d-quark- and the
electron Yukawa couplings are also perturbatively forbidden, we get
45 realistic models. For 11 of these 45 models, a large fine-tuning is required to
avoid large family mixing and to obtain neutrino masses in the desired range.  Note that  all setups for which the left-handed quarks $q_L$ arise from two sectors, namely $ab$ and $ab'$, suffer under a too large mixing between different families. To overcome such a mixing a large amount of fine-tuning is required. Nevertheless we display all 45 solutions in Table \ref{table four stack model  1/6a + 1/2 c -1/2 d}. We indicate the 11 solutions which generically give rise to a unrealistic CKM matrix with a $^{\clubsuit}$.
Of these 45 solutions, there are 18 models in which the $\mu$-term is perturbatively forbidden and gets generated
by an instanton with appropriate charge. In order to get a $\mu$-term of the desired order $(10^2-10^3) GeV$
the suppression factor is expected to be in the range $10^{-16}-10^{-15}$ . In 2 of these quivers, the instanton that is needed to generate the $\mu$-term would also induce the R-parity violating coupling $q_L L d_R$. Since the suppression factor of the $\mu$ term generating instanton is highly suppressed we do not exclude these setups. They are marked  with a $^\heartsuit$.

Furthermore, 14 quivers may give rise to additional massless $U(1)$'s which forbid some of the desired Yukawa couplings. All these are indicated with a $\dagger$. Again, if the potential massless $U(1)$'s become massive via the Green-Schwarz mechanism, then these quivers exhibit realistic Yukawa textures. If massless, though, the additional $U(1)$'s might prevent the generation of some desired Yukawa couplings.

Finally, 3 of these quivers run into the same issue discussed in section \ref{sec threestack1 }, where the instanton which generates the Dirac neutrino mass term $L H_u N_R$ also generates the Majorana mass term $N_R N_R$. In such a case, the seesaw masses are far below the experimentally observed order. Again we mark these quivers with a $^{\bullet}$.

\begin{table}[h]
\hspace{-.2cm}
\scalebox{.55}{
\begin{tabular}{|c|c|c|c|c|c|c|c|c|c|c
		|c|c|c|c|c|c|c|c|c|c|c|c|}\hline
		\multirow{2}{*}{Solution \#}&\multicolumn{2}{|c}{$q_L$} & \multicolumn{3}{|c}{$d_R$} & \multicolumn{2}{|c}{$u_R$} & \multicolumn{3}{|c}{$L$} & \multicolumn{3}{|c}{$E_R$}
		& \multicolumn{4}{|c}{$N_R$} & \multicolumn{4}{|c}{$H_u$} & \multicolumn{1}{|c|}{$H_d$}\\ \cline{2-23}
		&$(a,b)$ & $(a,\ov{b})$ & $(\ov{a},c)$ & $(\ov{a},\ov{d})$ & ${\Yasymm}_a$ & $(\ov{a},\ov{c})$
		& $(\ov{a},d)$ & $(b,\ov{c})$  & $(b,d)$ & $(\ov{b},d)$ & $(c,\ov{d})$
		& ${\Ysymm}_c $ & $\ov{\Ysymm}_d $ & $\Yasymm_b $ & $\ov{\Yasymm}_b $ & $(c,d) $ & $(\ov{c},\ov{d}) $
		& $(b,c)$ & $(\ov{b},c)$ & $(b,\ov{d})$ & $(\ov{b},\ov{d}) $ & $(\ov{b},\ov{c})$ \\
		\hline\hline			
		% 1&3&0&3&0&0&0&3&0&0&0&3&0&0&3&2&0&0&1&0&0&0&1&1\\ \hline did this one by hand
1$^{\dagger}$&3&0&3&0&0&0&3&0&0&3&0&0&3&2&0&0&1&0&0&0&1&1\\ \hline
2&3&0&3&0&0&0&3&0&0&3&1&0&2&2&0&1&0&0&0&0&1&1\\ \hline
3$^{\dagger}$&3&0&2&0&1&0&3&0&0&3&0&0&3&2&0&0&1&0&0&0&1&1\\ \hline
4&3&0&2&0&1&0&3&0&0&3&1&0&2&2&0&1&0&0&0&0&1&1\\ \hline
5$^{\dagger}$&3&0&1&0&2&0&3&0&0&3&0&0&3&2&0&0&1&0&0&0&1&1\\ \hline
6&3&0&1&0&2&0&3&0&0&3&1&0&2&2&0&1&0&0&0&0&1&1\\ \hline
7$^{\dagger}$&3&0&0&0&3&0&3&0&0&3&0&0&3&2&0&0&1&0&0&0&1&1\\ \hline
8$^{\dagger}$&3&0&0&0&3&0&3&0&0&3&0&0&3&3&0&0&0&1&0&0&0&1\\ \hline
9&3&0&0&0&3&0&3&0&0&3&1&0&2&2&0&1&0&0&0&0&1&1\\ \hline
10&3&0&3&0&0&2&1&0&0&3&2&1&0&2&0&1&0&0&0&0&1&1\\ \hline
11&3&0&3&0&0&2&1&0&0&3&0&2&1&2&0&1&0&0&0&0&1&1\\ \hline
12&3&0&3&0&0&3&0&0&0&3&2&1&0&2&0&0&1&0&1&0&0&1\\ \hline
13&3&0&3&0&0&3&0&0&0&3&0&2&1&2&0&0&1&0&1&0&0&1\\ \hline
14&3&0&3&0&0&3&0&0&0&3&1&2&0&2&0&1&0&0&1&0&0&1\\ \hline
15$^{\clubsuit}$&2&1&3&0&0&1&2&0&0&3&0&0&3&0&0&0&3&1&0&0&0&1\\ \hline
16$^{\clubsuit\,\bullet}$&2&1&3&0&0&1&2&0&0&3&3&0&0&0&0&3&0&1&0&0&0&1\\ \hline
17$^{\clubsuit\,\bullet}$&2&1&3&0&0&1&2&0&0&3&1&1&1&0&0&3&0&1&0&0&0&1\\ \hline
18$^{\clubsuit}$$^{\dagger}$&2&1&3&0&0&3&0&0&0&3&3&0&0&0&0&0&3&1&0&0&0&1\\ \hline
19$^{\clubsuit}$&2&1&3&0&0&3&0&0&0&3&1&1&1&0&0&0&3&1&0&0&0&1\\ \hline
20$^{\clubsuit\,\bullet}$&2&1&3&0&0&3&0&0&0&3&0&3&0&0&0&3&0&1&0&0&0&1\\ \hline
21$^{\clubsuit}$&2&1&3&0&0&3&0&0&2&1&2&1&0&2&0&0&1&1&0&0&0&1\\ \hline
22$^{\clubsuit}$&2&1&3&0&0&3&0&0&2&1&1&2&0&2&0&1&0&1&0&0&0&1\\ \hline
23$^{\clubsuit}$&1&2&3&0&0&3&0&0&1&2&2&1&0&0&2&0&1&1&0&0&0&1\\ \hline
24$^{\clubsuit}$$^{\dagger}$&1&2&3&0&0&3&0&0&3&0&3&0&0&0&0&0&3&1&0&0&0&1\\ \hline
25$^{\clubsuit}$&1&2&3&0&0&3&0&0&3&0&1&1&1&0&0&0&3&1&0&0&0&1\\ \hline
26$^\heartsuit$&0&3&0&3&0&0&3&3&0&0&2&0&1&0&3&0&0&0&0&1&0&1\\ \hline
27$^\heartsuit$&0&3&0&3&0&0&3&3&0&0&0&1&2&0&3&0&0&0&0&1&0&1\\ \hline
28$^{\dagger}$&0&3&0&0&3&0&3&0&3&0&0&0&3&0&3&0&0&1&0&0&0&1\\ \hline
29$^{\dagger}$&0&3&0&0&3&0&3&1&2&0&1&0&2&0&3&0&0&0&0&1&0&1\\ \hline
30&0&3&0&0&3&0&3&3&0&0&2&0&1&0&3&0&0&0&0&1&0&1\\ \hline
31&0&3&0&0&3&0&3&3&0&0&0&1&2&0&3&0&0&0&0&1&0&1\\ \hline
32&0&3&0&3&0&1&2&3&0&0&3&0&0&0&3&0&0&1&0&0&0&1\\ \hline
33&0&3&0&3&0&1&2&3&0&0&1&1&1&0&3&0&0&1&0&0&0&1\\ \hline
34$^{\dagger}$&0&3&0&0&3&1&2&1&2&0&2&0&1&0&3&0&0&1&0&0&0&1\\ \hline
35$^{\dagger}$&0&3&0&0&3&1&2&1&2&0&0&1&2&0&3&0&0&1&0&0&0&1\\ \hline
36&0&3&0&0&3&1&2&3&0&0&3&0&0&0&3&0&0&1&0&0&0&1\\ \hline
37&0&3&0&0&3&1&2&3&0&0&1&1&1&0&3&0&0&1&0&0&0&1\\ \hline
38&0&3&0&0&3&2&1&0&3&0&3&0&0&0&3&0&0&1&0&0&0&1\\ \hline
39&0&3&0&0&3&2&1&0&3&0&1&1&1&0&3&0&0&1&0&0&0&1\\ \hline
40$^{\dagger}$&0&3&0&0&3&2&1&2&1&0&2&1&0&0&3&0&0&1&0&0&0&1\\ \hline
41$^{\dagger}$&0&3&0&0&3&2&1&2&1&0&0&2&1&0&3&0&0&1&0&0&0&1\\ \hline
42$^{\dagger}$&0&3&0&3&0&3&0&3&0&0&0&3&0&0&3&0&0&1&0&0&0&1\\ \hline
43$^{\dagger}$&0&3&0&2&1&3&0&3&0&0&0&3&0&0&3&0&0&1&0&0&0&1\\ \hline
44$^{\dagger}$&0&3&0&1&2&3&0&3&0&0&0&3&0&0&3&0&0&1&0&0&0&1\\ \hline
45&0&3&0&0&3&3&0&1&2&0&1&2&0&0&3&0&0&1&0&0&0&1\\ \hline
\end{tabular}}
\caption{Spectrum for the solutions with $U(1)_Y= \frac{1}{6} U(1)_a +\frac{1}{2} U(1)_c -\frac{1}{2}U(1)_d $.
\label{table four stack model  1/6a + 1/2 c -1/2 d}
}
\end{table}

\subsection{$SU(2)$ Realized as $Sp(2)$
\label{chapsp(2)}
}

As discussed previously, a stack of $N$ coincident D-branes in general gives rise to a
$U(N)$ gauge group. If the stack wraps an orientifold invariant cycle, however, then the
gauge group is $Sp(2N)$. Since $Sp(2)$ is isomorphic to $SU(2)$, we can realize the $SU(2)_L$
of the MSSM as a $U(2)$ from a D-brane stack on a generic cycle or as an $Sp(2)$ arising from a
D-brane stack wrapping an orientifold invariant cycle. The former was the subject in sections
\ref{sec three-stack models} and \ref{sec four-stack models}, and here we focus on the latter.
In this case, all representations are real and it is easy to see that the tadpole equations
do not impose any condition on the transformation behavior under the $Sp(2)$. This suggests
that there might be more solutions than for the $U(2)$ realization of $SU(2)_L$ considered
in the previous chapter.

This conclusion turns out to be too naive for two reasons. First, the $Sp(2)$
does not exhibit a $U(1)_b$ which could contribute to the hypercharge.
This restricts the possible number of hypercharge choices to the subset of
the previously analyzed ones which do not have any contribution from $U(1)_b$, namely
\begin{itemize}
\item[$\bullet$] $U(1)_Y=\frac{1}{6} U(1)_a +\frac{1}{2} U(1)_c$
\item[$\bullet$]$U(1)_Y= \frac{1}{6} U(1)_a +\frac{1}{2} U(1)_c - \frac{3}{2}U(1)_d$
\item[$\bullet$]$U(1)_Y= \frac{1}{6} U(1)_a +\frac{1}{2} U(1)_c -\frac{1}{2}U(1)_d$.
\end{itemize}
Here the first choice can be realized as three or four stack model while the
latter two are quivers based on four stacks of D-branes.

The second reason why there are only a few MSSM realizations is due to the
fact that the stack $b$ is identified with its image stack $b'$. This limits the potential
origins of fields charged under the $SU(2)_L$. For instance, for the first two hypercharges
the leptons and the Higgs pair arise from the same sector $bc$. Thus for these setups the $\mu$-term
is perturbatively realized, but also the R-parity violating coupling $L\,H_u$ is present, which make
these configurations unrealistic.

This leaves us with the hypercharge $U(1)_Y= \frac{1}{6} U(1)_a +\frac{1}{2} U(1)_c -\frac{1}{2}U(1)_d$, which we now discuss.

\subsubsection{$U(1)_Y= \frac{1}{6} U(1)_a +\frac{1}{2} U(1)_c -\frac{1}{2}U(1)_d$}

\begin{align*}
q_L\,:&\qquad (a,b)\\
u_R\,:&\qquad (\ov a, \ov c),\,\,\,\,(\ov a,  d)\\
d_R\,:&\qquad (\ov a,  c),\,\,\,\,     (\ov a, \ov d), \,\,\,\, \Yasymm_a\\
L\,\,:&\qquad (b,\ov c),\,\,\,\,(b, d)\\
E_R\,:&\qquad (c,\ov d),\,\,\,\, \Ysymm_c,\,\,\,\,\,  \ov \Ysymm_d\\
N_R\,:&\qquad (c,d),\,\,\,\, (\ov c, \ov d)\\
H_u\,:&\qquad (b,c),\,\,\,\,(b,\ov d)\\
H_d\,:&\qquad (b,\ov c),\,\,\,\,(b, d)
\end{align*}

For this choice of massless hypercharge, we obtain 100 tadpole free models, 8 of which don't give rise to R-parity violating couplings. Taking into account non-perturbative effects, we are left with one setup, which also has a perturbatively present top Yukawa coupling.
This setup is presented in Table \ref{table spectrum four stack model Sp2
1/6a+1/2c-1/2d}.
\begin{table}[h]
\centering
\begin{tabular}{|c|c|c|c|c|c|c|c|c|}\hline
		\multirow{2}{*}{Solution \#}&$Q_L$&$d_R$&$u_R$&$L$&$E_R$&$N_R$&$H_u$&$H_d$\\
        \cline{2-9}
		&$(a,b)$&$(\ov{a},c)$&$(\ov{a},\ov{c})$&$(b,d)$&$(c,\ov{d})$&$(\ov{c},\ov{d})$&$(b,c)$&$(b,\ov c)$\\
		\hline\hline			
		1$^{\dagger}$&3&3&3&3&3&3&1&1\\ \hline		
\end{tabular}
\caption{Spectrum for the solution with $U(1)_Y=\frac{1}{6}\,U(1)_a+\frac{1}{2}\,U(1)_c-\frac{1}{2} U(1)_d$.
\label{table spectrum four stack model Sp2
1/6a+1/2c-1/2d}}
\end{table}

This quiver, whose local realization has been studied in \cite{Cremades:2002va,Cremades:2003qj}, exhibits all
 desired Yukawa couplings on the perturbative level
\begin{align} \nonumber
< {q^I_L}_{(1,0,0)}\, {H_u}_{(0,1,0)}\, {u^J_R}_{(-1,-1,0)} > \qquad < {q^I_L}_{(1,0,0)}\, {H_d}_{(0,-1,0)}\, {d^J_R}_{(-1,1,0)} >\\
<L^I_{(0,0,1)}\,{H_d}_{(0,-1,0)}\, {E^J_R}_{(0,1,-1)}> \qquad  <L^I_{(0,0,1)}\,{H_u}_{(0,1,0)}\, {N^J_R}_{(0,-1,-1)}>\\
<{H_u}_{(0,1,0)} \,{H_d}_{(0,-1,0)}>\,\,.\qquad \qquad \qquad \qquad \qquad \nonumber
\end{align}
Therefore, the mass hierarchy between the quarks and leptons, as well as the hierarchies
within the families, are due to the worldsheet instantons rather than spacetime instantons.

Since the Dirac neutrino masses are perturbatively realized, they are expected to be of the same order as the masses for the other leptons. As in \ref{sec threestack1 }, a Majorana mass term of the right order can account for the measured smallness of neutrino masses via the type I seesaw mechanism. Such a term can be induced by an instanton with the intersection pattern
\begin{align}
I_{E2a}=0  \qquad I_{E2b}=0 \qquad I_{E2c}=-2 \qquad I_{E2d}=-2,
\end{align}
where the suppression factor should be in the range $10^{-5}$ to $10^{-2}$ to account
for the observed neutrino masses. Note that this instanton does not induce any undesired R-parity violating couplings.
This setup may gives rise to an additional massless U(1)
\begin{align}
U(1)^{add}=U(1)_c\,\,,
\end{align}
which if indeed present would forbid the Majorana mass term for the right-handed neutrino.

\section{Conclusion \label{sec conclusion}}

In this work we systematically investigate MSSM D-brane quivers, arising from three and four stacks of D-branes, with respect to their Yukawa structure.
For almost all quivers, various desired Yukawa couplings are perturbatively forbidden due to global $U(1)$ selection rules.
D-brane instanton effects can generate these missing couplings and may also account for various hierarchies.
Here we analyzed the implications of such non-perturbative effects for the phenomenology of the respective
quivers. We find that often times the desired Yukawa coupling inducing instanton also leads to phenomenologically undesired
effects. The latter include the generation of R-parity violating couplings, of $N_R$ tadpoles, of a
$\mu$-term which is too large, or too large family mixing. Subsequently, such quivers are ruled out as unrealistic.

Furthermore, we require that a viable quiver exhibits a mechanism which can account for the smallness of the neutrino masses.
In this work we considered two different scenarios.
The first scenario is the well known type I seesaw mechanism, which requires the presence of a large Majorana mass
term for the right-handed neutrinos. Such a mass term is perturbatively forbidden,
but can be induced by D-instantons with the right zero mode structure.
The second scenario assumes that the Dirac mass term violates global $U(1)$
selection rules and thus is perturbatively forbidden. A D-instanton with high
suppression factor which compensates for the global U(1) charge carried
by the Dirac mass term can account for the small neutrino masses.

In section \ref{sec three-stack models} we analyze in detail D-brane quiver
with respect to these constraints. We find only one realistic quiver, displayed
in Table \ref{spectrum three stack model 1.1}. All others suffer from some phenomenological drawbacks.
Either the Yukawa coupling inducing instanton generates R-parity violating couplings,
or the top Yukawa coupling is perturbatively forbidden, while some d-quark couplings are perturbatively realized.
The latter is in contradiction with observations, which suggest the opposite hierarchy.
We also encounter one quiver that does not allow for neutrino masses in the observed range.
This is due to the fact that the Dirac mass generating instanton also induces the Majorana mass
term for the right-handed neutrinos. Thus the  seesaw mass is effectively a non-perturbatively
generated Weinberg operator which generically gives too small neutrino masses \cite{Ibanez:2007rs}.

Equipped with what we learned from the three stack quivers, we perform a systematic search for phenomenologically
viable D-brane quivers based on four D-brane stacks. We show that only a small subset of the
D-brane quivers that satisfy the two top-down constraints, tadpole cancellation and presence of a
massless hypercharge, give rise to phenomenology compatible with experimental observations.
We display these quivers in the tables \ref{table spectrum four stack model 1/6a+1/2c-3/2d},
 \ref{table four stack model  1/3a - 1/2 c}, \ref{table four stack model  1/3a - 1/2 b+1 d},
\ref{table four stack model  1/6a + 1/2 c -1/2 d} and \ref{table spectrum four stack model Sp2
1/6a+1/2c-1/2d}. These quivers serve a starting point for future quests for concrete MSSM realizations with realistic Yukawa texture.

Some of these quivers potentially exhibit additional massless $U(1)$'s. These additional symmetries have to be preserved by the superpotential and thus often times various desired Yukawa couplings are forbidden, even at the non-perturbative level. However, the conditions derived in section \ref{sec massless U(1)}, which are constraints on the transformation behavior of the matter fields arising from the masslessness condition, are only necessary conditions. The actual masslessness condition is a constraint on the cycles. Thus, it needs to checked for a concrete realization if such additional $U(1)$'s are indeed present.

Some quivers discussed in sections \ref{sec three-stack models} and
\ref{sec four-stack models} may exhibit dimension
five operators which could lead to rapid proton decay, unless they are
sufficiently suppressed. We leave it for future work
\cite{cvetic:2009new} to analyze the constraints arising from the
considerations of these operators (for a similar analysis for $ SU(5)$
orientifolds, see \cite{Kiritsis:2009sf}).

This bottom-up analysis shows that family splitting, namely that different families of the same matter field arise from different sectors is phenomenologically disfavored. This splitting has been used in \cite{Anastasopoulos:2009mr} as a mechanism to explain the different mass hierarchies in the MSSM. We expect that increasing the number of D-brane stacks allows for MSSM quivers which exhibit such family splitting while not containing phenomenological drawbacks, thus giving a mechanism to explain observed mass hierarchies. It would be interesting to extend the current analysis by performing a detailed analysis of Yukawa textures for higher-stack quivers \cite{cvetic:2009new}.

We would also like to point out that the approach presented here has broader
applications to other corners of the string landscape.
While the concrete analysis has been carried out explicitly in the Type
IIA context, it has a straightforward map to Type I constructions  with
magnetized D9-branes. Analogous studies can be carried
out along the same lines in the Type IIB context with D-branes at
singularities. In such a case, however, the analysis of the corresponding  quivers
is carried out in a geometric regime of closed sector moduli which
are T-dual to a  non-geometric regime of Type IIA analysis (and vice
versa).

In this work we restrict the spectrum to be the exact MSSM spectrum plus three right-handed neutrinos.
As shown in \cite{Ibanez:2008my}, allowing for an additional Higgs pair can overcome the problem of a too large $\mu$-term.
We leave it for future work to systematically analyze such D-brane quivers. Furthermore, one might entertain the idea
of allowing additional singlets under the standard model gauge groups, which could acquire a VEV and then induce some of the desired Yukawa couplings \cite{Anastasopoulos:2009mr} via higher order couplings.  It would be interesting to see if the splitting of standard model families is still phenomenologically disfavored in an analysis which allows for additional chiral singlets.

\vspace{2cm}

{\bf Acknowledgments}\\
We thank T. Brelidze, I. Garc\'ia-Etxebarria, E. Kiritsis, A. Lionetto, P. Langacker, S. Raby, B. Schellekens and T. Weigand for
useful discussions. We are especially grateful to L. E. Ib{\'a}{\~n}ez for
intensive discussions and correspondence in the early stage of the project.
The work is supported by NSF RTG grant DMS-0636606 and  Fay R. and Eugene L. Langberg
Chair.
\clearpage \nocite{*}
\bibliography{rev}
\bibliographystyle{utphys}

\end{document}